\documentclass[11pt]{article}
\usepackage[margin=1in]{geometry}
\usepackage{amsmath,amssymb}
\usepackage[nopatch]{microtype}
\usepackage{booktabs}
\usepackage{threeparttable}
\usepackage{graphicx}
\usepackage{csquotes}
\DeclareMathOperator{\cov}{cov}
\DeclareMathOperator{\var}{var}
\usepackage[backend=biber,style=apa]{biblatex}

\usepackage[T1]{fontenc}
\usepackage{lmodern}
\usepackage{booktabs,longtable,array}
\usepackage{setspace}
\usepackage{pdflscape}
\usepackage{url}
\usepackage{makecell}
\usepackage{float}
\usepackage{placeins}
\usepackage{caption}

\usepackage{xcolor}
\usepackage{hyperref}
\hypersetup{
    colorlinks=true,       % true: colored links; false: boxed links
    linkcolor=blue,        % color of internal links
    citecolor=blue,        % color of links to bibliography (citations)
    filecolor=blue,        % color of file links
    urlcolor=blue          % color of external links (URLs)
}

\title{\bf Shift-Share Designs in Political Science\thanks{Peter Park tragically passed away in December 2025. As his advisors, we assembled this manuscript and posted it on arXiv to preserve and share his work. The most meaningful way to remember Peter is to carry forward his intellectual legacy. The manuscript is not fully complete; I have made only minimal formatting edits. Yiqing Xu. Email: \url{yiqingxu@stanford.edu}.
}}

\author{Peter Kyungtae Park\\
Department of Political Science, Stanford University\\
\texttt{kyungtae.park@stanford.edu}}

\date{September 17, 2025}

\begin{document}

\maketitle

\begin{abstract}\noindent 
Shift-share designs are gaining popularity in political science. This article introduces what shift-share designs are, reviews their application in the literature, synthesizes recent methodological developments, and discusses their potential utility in the field. Although shift-share designs have a long historical use in economics, their causal properties only recently began to be understood. Articles in political science tend to be aware of these developments, but do not fully discuss and test identifying assumptions and sometimes apply the methods incorrectly. Most articles rely on the share exogeneity framework, suggesting that the shifter exogeneity framework is underutilized despite its comparable prevalence in economics. I illustrate shifter exogeneity framework and develop auxiliary theoretical results that are potentially useful in applying the framework in political science settings.\\ \medskip

\end{abstract}

\thispagestyle{empty}
\clearpage

\section{Introduction}
\label{sec:introduction}
A comprehensive introduction to shift-share methods lacks despite a growing interest within political
science. The need arises partly from the complexity of the methods themselves, and partly from that of
the literature. The historical literature can be confusing without backgrounds as the term shift-share has
appeared in various contexts. Moreover, recent methodological developments have yet to be organically
synthesized for those unfamiliar with the methods. \footnote{\textcite{BorusyakHullJaravel2025b} synthesize the literature from a methodological viewpoint but assume some familiarity
with shit-share methods as they are prevalent in modern economics. This article can be read together with theirs.\textcite{BorusyakHullJaravel2025a} a technical review.} This article aims to fill both gaps by introducing
the usage of the methods in the literature and the new identification strategies, while maintaining their
relevance for political scientists.

The first half of the article reviews the economics literature. Shift-share designs refer to research de-
signs with shift-share measures. Suppose $n$ units are subject to $m$ common treatments by varying degrees.
Formally, unit $i$ is exposed to the $j$th treatment $D_j$ (``shift'') by a non-negative
weight $w_{ij}$ (``share''), where weights sum to less than $1$ in each unit: $w_{ij} \ge 0
\quad \text{and} \quad
\sum_{j=1}^{m} w_{ij} \le 1$. 
The shift-share measure for unit $i$ is defined as
\begin{equation}
X_i = \sum_{j=1}^{m} w_{ij} D_j .
\label{eq_1}
\end{equation}

We use the China shock for the working example. Other leading examples are in Section~\ref{sec:examples}.
\textcite{AutorDornHansonMajlesi2020} study whether imports from China caused political polarization in the United States. Regions are exposed to common national trade shocks, but by varying degrees depending on the presence
of each industry in the regions. Under the above notation, units $i$ are regions, treatment units $j$
are industries, and weights $w_{ij}$ are regional employment shares of each industry. Shifts $D_j$
compute the decadal national change in Chinese imports divided by the national demand for each industry,
referred to as import penetration.

The shift-share measure performs two tasks. First, it provides a summary measure for comparable yet potentially heterogeneous shifts. While different industries may have different effects on regional electoral outcomes, a single regional import shock measure may reasonably summarize these varied effects. Second, it imputes an unobservable unit-level treatment with proxy shifts. A direct measure of the import shock would exploit the \textit{regional} import change divided by the \textit{regional} demand for each industry. Since trade data is often only available at the national level, the shift-share measure proxies the direct one by combining national import data with regionally available data.

However, employing a measure that combines two sets of variables complicates the statistical analysis. Establishing the exogeneity of even a single source is already a demanding task that warrants a dedicated section in most empirical articles. For example, shifts might not be exogenous to the outcome variable in our working example. Consider two regions that have similar industry portfolios. They are likely to share similar demographic composition and economic interests, and hence prone to similar unobserved electoral shocks. These confounders may lead to the omitted variable bias if unaccounted for. The concerns about identification have emerged only more recently despite the long history of the use of the methods.

The literature thus proceeded with the exogeneity of either shifts or shares instead of both. The share exogeneity framework assumes that shares are exogeneous conditional on covariates \parencite{GoldsmithPinkhamSorkinSwift2020}. This framework allows shifts to be determined endogenously to unit-level outcomes,\footnote{Since shifts are common across units, such endogeneity may exist with respect to the joint distribution of the unit-level
outcomes --- \textit{e.g.} when shifts covary with the average of the stochastic terms in the outcome variable.} yet the relative differences in shifts effectively applied across units are exogenously determined through the shares. This setting is a classical difference-in-differences design if the multiple shifts were considered separately.

The shift exogeneity framework assumes that shifts are exogeneous conditional on covariates \parencite{AdaoKolesarMorales2019,BorusyakHullJaravel2022}. The difference-in-differences analogy breaks down since the relative differences are now determined by endogenous shares. One might instead hope that if no single share is too large, the biases due to the endogeneity in the relative differences cancel out much like the law of large numbers. Such cancellation requires not only small shares but also a large number of shifts. \textcite{BorusyakHull2023} propose an alternative method using randomized inference under a stronger assumption that shifts follow the same distribution and are interchangeable.

The second half of this article reviews the use of shift-share designs in political science. Shift-share designs are increasingly common, especially in areas close to economics such as technology shock, capital movement and foreign aid. Many studies directly import shift-share variables developed in economics into political science applications. While these studies have generated important findings in their own right, the methods appear to hold even greater potential within political science. It is especially striking that American politics has seen the fewest applications of shift-share methods given the field’s unparalleled access to rich data on complex interactions among different sets of actors, including elections and donation networks.

However, identification is not often justified thoroughly even in recently published articles. Only about half of the articles published since 2021 cite one of the above two exogeneity frameworks. Moreover,Many of those that do merely state the assumptions without substantively arguing for them or conducting falsification tests to assess their plausibility. The fact that articles mostly rely on share exogeneity may suggest either that the framework is being misapplied or that the shift exogeneity framework remains underutilized. I propose three dimensions of shift-share designs to consider when designing shift-share variables and their identification strategies.

I replicate \textcite{ColantoneStanig2018b} to illustrate how to discuss identifying assumptions. The authors study the impact of Chinese import on electoral outcomes, similarly to \textcite{AutorDornHansonMajlesi2020} but in European countries instead of the United States. \textcite{BorusyakHullJaravel2022} argue that the original China shock measures by \textcite{AutorDornHanson2013} that use the US regional employment shares align with the shift exogeneity framework. This article finds that their European counterpart needs an additional shift transformation to justify the framework, and that correct procedures overturn the statistical significance of the findings. This exercise demonstrates that researchers must take extra care to tailor shift-share designs to their specific contexts. The transformation and residualization schemes developed here can be applied in other shift-share designs. talk more about residualization.

The paper is organized as follows. Section~\ref{sec:examples} introduces three leading examples that further motivate shift-share designs. Section~\ref{sec:history} surveys the historical use of shift-share designs in economics. Section~\ref{sec:identification} synthesizes two main identification strategies under the share exogeneity and the shift exogeneity. Section~\ref{sec:ps} reviews the use of shift-share designs in political science and makes comments. Section~\ref{sec:china} illustrates the shift exogeneity framework through replicating a paper. Section~\ref{sec:conclusion} concludes the paper.

\section{Motivating examples}
\label{sec:examples}

\textcite{AutorDornHansonMajlesi2020} Continuing from Section~\ref{sec:introduction}, both shares and shifts can be endogenous to electoral outcomes if unobserved domestic confounders affect employment, import and politics at the same time. \footnote{\textcite{AdaoKolesarMorales2019} a structural model that induces non-trivial correlations between employment, import
and wage in the context of \textcite{AutorDornHanson2013}} The authors construct a shift-share instrument that interacts local US employment with Chinese import to other advanced economies comparable to the US:
\[
Z_i = \sum_j w_{ij} D_j^{*},
\]
where $D_j^{*}$ denotes the counterfactual shifts. This instrument isolates the export shock originating from China, which is plausibly exogenous to US electoral outcomes. The implied identification strategy is shift exogeneity.

\textcite{FoukaTabellini2022} The authors ask whether Mexican immigrants to the United States have changed white Americans’ perception of Black Americans. Since immigrants can endogenously choose where to move in upon observing local characteristics, they construct a ``Bartik'' instrument that interacts the initial regional share of Mexicans $w_i$ with the national inflow of Mexicans in subsequent decades $D_t$: 
$Z_{it} = w_i D_t$.\footnote{$Z_{it}$ is then scaled by the predicted regional population. The methodological implications will be discussed in Section~\ref{sec:ps}.} The instrument redistributes the national immigrant inflow accordingly to the initial distribution of immigrants. The national inflow may correlate with the joint distribution of regional perceptions (see footnote 2), but the initial regional share $w_i$ is free from any time-varying confounders. The identification strategy is share exogeneity.

\textcite{NunnQian2014} The authors asks how US food aid affects the incidence of conflict in recipient countries. Since food aid may be triggered by conflict, they instrument aid with US wheat production, noting that wheat aid has been used as a way to institutionally handle surplus food production. To maximize the instrument power, they further interact wheat production with the fraction of years a recipient country received food aid during the study period: $Z_{it} = w_i D_t$ 
where $w_i$ is the fraction and $D_t$ is wheat production. The instrument captures the prospective amount of food aid determined by a potentially endogenous history $w_i$ (referred to as ``propensity score'' in some articles) and an exogenous food production $D_t$. The implied identification strategy is shift exogeneity.

\section{Brief overview of shift-share methods}
\label{sec:history}
This section traces the historical development of shift-share methods and shows their wide application in the economics literature. The term ``shift-share'' was first coined in the context of shift-share analysis or shift-share decomposition, a technique that descriptively decomposes a single variable into several components \parencite{Dunn1960,EstebanMarquillas1972,Lemieux2002}. As one of the first examples, \textcite{Perloff1957} calculates the ``expected'' income per capita of each state and the deviance from the expected income using the national average income and the local industry shares. Denote $D_{ij}$ for the income per capita of state $i$ and industry $j$, and $w_{ij}$ for the employment share of industry $j$ in state $i$. The income per capita $X_i$ of state $i$ can be decomposed into
\[
X_i
= \sum_j w_{ij} D_{ij}
= \underbrace{\sum_j w_{ij} \bar D_{\cdot j}}_{\text{expected income}}
+ \underbrace{\sum_j w_{ij}\left(D_{ij}-\bar D_{\cdot j}\right)}_{\text{regional income shock}} ,
\]

where $\bar D_{\cdot j}$ is the national average income of industry $j$.

$Z_{it}$ is then scaled by the predicted regional population. The methodological implications will be discussed in Section~\ref{sec:ps}.

If the share $w_{ij}$ varies over time so $w_{ijt}$ denotes the share $w_{ij}$ at time $t=0,1,\ldots$, we can express $w_{ijt}$ as the sum of the initial share and the change over time $w_{ij0} + (w_{ijt}-w_{ij0})$ and break down $X_i$ into three components:
\begin{equation}
X_{it}
= \sum_{j} w_{ijt} D_{ijt}
= \underbrace{\sum_{j} w_{ij0}\bar D_{\cdot jt}}_{\text{expected income}}
+ \underbrace{\sum_{j} w_{ij0}\bigl(D_{ijt}-\bar D_{\cdot jt}\bigr)}_{\text{regional income shock}}
+ \underbrace{\sum_{j} \bigl(w_{ijt}-w_{ij0}\bigr)\bar D_{\cdot jt}}_{\text{change in shares}} .
\tag{2}
\label{eq:shiftshare-decomp}
\end{equation}

Equation~\eqref{eq:shiftshare-decomp} has two deviance terms each due to the change in shares and to idiosyncratic income shocks. This technique is often used to assess the relative explanatory power of multiple possible causes of phenomena \parencite{BoundJohnson1992,OlivettiPetrongolo2016,BursteinMoralesVogel2019,FreemanGanguliHandel2020}. For example, Equation~\eqref{eq:shiftshare-decomp} can shed light on whether industry-level income shocks or changes in shares contributed more to the average change in income. While modern literature has advanced into non-linear or nonparametric decompositions, this linear method remains popular for its simplicity.

Shift-share methods later appeared in structural models as Bartik instruments. Bartik (1991) regresses local wage on local labor supply to estimate the inverse elasticity of labor supply. Since price and quantity are simultaneously determined in the market, the author introduces an instrument inspired by equation (2). Redefining $D_{ijt}$ as labor supply of region $i$, industry $j$ and time $t$, the idea is that the expected labor supply $\sum_j w_{ij0}\bar D_{\cdot jt}$ is correlated with the actual labor supply $X_{it}$ through the initial share $w_{ij0}$ but not with time-varying confounders. The argument relies on the assumption that national labor supply $\bar D_{\cdot jt}$ reflects labor demand shocks rather than labor supply shocks, leaving open the critique noted in footnote~2.

This instrumentation differs from shift-share decomposition in that the dimension $j$ is not predefined. Researchers are free to choose any dimension $j$ to construct Bartik instruments as long as the independent variable $X_i$ can be decomposed along the dimension $j$ and a plausible ``average'' $\bar D_{\cdot j}$ exists. Therefore, multiple instruments may exist for the same independent variable depending on how it is decomposed (the dimension of $j$) and how individual shifts are approximated (the nature of $\bar D_{\cdot j}$).

Bartik instruments subsequently gained popularity in the wider economics literature. As region and industry are two natural orthogonal levels of analysis, economists have constructed Bartik instruments by interacting the same industry shares with various national outcomes such as earnings \parencite{Luttmer2005, diamond2016location}, hours worked \parencite{BoundJohnson1992, BoundHolzer2000}, productivity and technology shocks \parencite{GouldWeinbergMustard2002, AcemogluRestrepo2020}, and immigration inflow \parencite{Card2001}. Regional variables can be decomposed along other dimensions such as immigration inflow by skill groups \parencite{Card2009}, income growth by income groups \parencite{BoustanFerreiraWinklerZolt2013}, credit shifts by banks \parencite{GreenstoneMasNguyen2020}, and price changes by housing characteristics \parencite{GrahamMakridis2023}.

Sometimes the term Bartik is used to emphasize the approximation of regional variables with national shifts when no decomposition is involved.  construct a counterfactual number of local Mexican immigrants using national immigrant inflows rather than decomposing the number of local immigrants. \textcite{GabrielKleinPessoaForthcoming} study the effect of austerity on political extremism and instrument regional austerity measures with national austerity measures, multiplied by the ratio of regional to national per capita government spending. The multiplier represents each region’s sensitivity to government spending.

Shift-share designs or variables focus on the special inner product structure of the variable as in Equation~\ref{eq_1}  with Bartik instruments as a special case. A prominent example is \textcite{AutorDornHanson2013} who first introduced the China shock measure and instrument. The authors use shift-share variables not only for instruments but also for noisy estimates of local exposure to Chinese imports. The ``China shock'' measure has drawn wide attention and has been applied to various macro outcomes such as employment \parencite{AcemogluAutorDornHansonPrice2016}, mortality \parencite{PierceSchott2020}, marriage rate \parencite{AutorDornHanson2019}, public goods provision \parencite{FelerSenses2017}, and polarization \parencite{AutorDornHansonMajlesi2020}; as well as to contexts outside the US \parencite{DauthFindeisenSuedekum2017,BaroneKreuter2021}.

Another difference between the China shock approach and Bartik instruments is that the former consider horizontal counterfactual shifts (Chinese export to the US versus to comparable economies) instead of hierarchical counterfactual shifts (regional shifts versus national average) to construct the instrument. This is to isolate the exogenous component of Chinese import attributable to Chinese factors rather than domestic ones, thereby making the shift exogeneity framework more appropriate. \textcite{StuenMobarakMaskus2012} study the effect of foreign doctoral student supply at US universities on scientific publications, and similarly build an instrument by interacting shares of students from each source country at a given university-field with the number of doctoral students choosing other host countries.

Other examples highlight the versatility of shift-share designs through creative choices of units and shares. \textcite{Kovak2013} measures regional exposure to trade liberalization by interacting price differentials per industry with composite shares derived from a structural model, which consist of regional industry labor shares, the elasticity of substitution between production factors, and their cost shares. \textcite{HummelsJorgensenMunchXiang2014} define firms as units and measure their exposure to transportation costs by interacting changes in national transportation cost with shares of input source countries for each firm. \textcite{AcemogluLinn2004} define new drug categories as units and measure their exposure to demographic changes by interacting demographic changes per age group and age profiles of users for each drug category. \textcite{XuChenzi2022} defines ports as units and measures their exposure to bank failure by interacting bank failure rates and their credit exposure to each bank.

The methodological definition of shift-share distills the common identification challenges that arise in any study using shift-share variables. While it does not prescribe how to construct shift-share measures or instruments compared to Bartik instruments, it conversely broadens applicability of the methods. Researchers are free to use any shift-share designs that satisfy the identifying assumptions.

\section{Identification and inference strategies}
\label{sec:identification}
This section summarizes identification and inference strategies in shift-share designs. \textcite{BorusyakHullJaravel2025b} provide high-level intuition for the two frameworks and practical guidance to commonly asked questions. This section produces a mathematical summary of the frameworks using consistent notations. This is to help potential users of the methods engage with the original articles that proposed these strategies from different motivations and perspectives. Appendix~B contains technical details.

The biggest challenge in shift-share designs is that shares and shifts are rarely both exogenous. One plausibly exogenous variation is already hard to find in observational studies; finding two is much harder. Two distinct statistical problems may arise. First, the identification problem: OLS or 2SLS estimates involving shift-share variables might be biased if units systematically select into different treatments due to non-randomness in shifts or shares. We will see that this does not happen under the identifying assumptions of either framework.

Second, the inferential problem: the typical cluster-robust standard error estimator may underestimate the true standard error even when the point estimate is valid. \textcite{AdaoKolesarMorales2019} show that this can occur when shares exhibit a non-trivial correlation structure that does not align with the chosen clustering scheme. Consider the China shock example. Errors are typically clustered at the state level in geographical studies. However, units with similar industry shares may have correlated errors regardless of geographical proximity since employment and wages---the outcome variable in \textcite{AutorDornHanson2013}---are functions of the same supply and demand shifters, potentially generating complex correlation structures. The authors therefore propose an inference procedure that accommodates arbitrary correlations in the error term. This inferential problem does not arise under share exogeneity.

\subsection{Share Exogeneity: Exogenous Shares}
\label{sec:4_1}

This section summarizes findings in \textcite{GoldsmithPinkhamSorkinSwift2020}Goldsmith-Pinkham. Share exogeneity holds when shares in the shift-share measure are orthogonal to the error term. We treat shifts as fixed to allow for an arbitrary correlation structure, and everything else as i.i.d.\ across units.

\paragraph{Identification} Consider a simple model with two shifts: 
$Y_i = \alpha + \beta X_i + \epsilon_i 
\quad \text{and} \quad
X_i = w_{i1}D_1 + w_{i2}D_2$.
The regression of $Y_i$ on $X_i$ yields
$\hat\beta = \frac{\widehat{\cov}(Y_i,X_i)}{\widehat{\var}(X_i)} 
= \beta + \frac{\widehat{\cov}(X_i,\epsilon_i)}{\widehat{\var}(X_i)}$,
and the covariance between $X_i$ and $\epsilon_i$ is
$\cov(w_{i1}D_1,\epsilon_i) + \cov(w_{i2}D_2,\epsilon_i)
= D_1\cdot\cov(w_{i1},\epsilon_i) + D_2\cdot\cov(w_{i2},\epsilon_i)$.

The last operation uses that $D_1$ and $D_2$ are fixed.\footnote{Fixing shifts means that if we allow shifts to be stochastic, consistency holds as long as the share exogeneity is satisfied for each realized value of shifts $(D_1,D_2)$. This of course would hold if shares were randomized.} Share exogeneity assumes that shares are uncorrelated with the error term: $\cov(w_{ij},\epsilon_i)=0$ for all $j$ such that $D_j\neq0$. Therefore, the OLS estimate $\hat\beta$ consistently estimates the true $\beta$ under the identifying assumptions.

Identification under share exogeneity is analogous to difference-in-differences designs. In a shift-share design with one shift, units are exposed to a single common shock but to varying degrees exogenously determined by shares. Share exogeneity thus ensures comparability between units. Designs with multiple shifts merely pool multiple difference-in-differences designs. This implies that neither the nature nor the number of shifts helps identification.

\paragraph{Inference} We can view the OLS regression as an IV regression where $X_i$ instruments itself. Consider an IV model where exogenous shares $w_{i1}$ and $w_{i2}$ instrument the shift-share measure $X_i$ separately:
\[
X_i = \gamma + \delta_1 w_{i1} + \delta_2 w_{i2} + \eta_i,
\]
\[
Y_i = \alpha + \beta X_i + \epsilon_i .
\]
Note that the model is identified as we use two instruments for one independent variable. The Generalized Method of Moments (GMM) estimator can impose arbitrary weights on the first-stage coefficients $\delta_1$ and $\delta_2$, but inference (and consistency) is valid regardless of the specific values at which weights are fixed. Weights $\delta_1/\delta_2=D_1/D_2$ immediately yield $\delta_1=D_1$, $\delta_2=D_2$ and the fitted first-stage value $\hat X_i=X_i$. This shows that shift-share regressions are a special case of GMM under share exogeneity.

\paragraph{Control variables} Zero covariance conditions may be weaker than mean-independence $\mathbb{E}[\epsilon_i \mid w_{ij}]=0$ or full independence $\epsilon_i \perp w_{ij}$, but they remains strong as they must hold for every shift $j$. We consider two relaxations. First, we allow shares to be exogenous conditionally on control variables:
$\cov(w_{ij},\epsilon_i \mid \Pi_i)=0$ where $\Pi_i$ denotes controls. These controls can be included directly in the OLS regression. To see why this works, observe that $\Pi_i$ can be included as exogenous variables in the IV regression.

\paragraph{Rotemberg decomposition} Second, we may instead require zero covariance conditions to hold only for the shifts that matter most for the final estimate. Let $\hat\beta_j$ denote the estimate instrumented with the $j$th share alone. The shift-share regression coefficient can then be expressed as a weighted average of $\hat\beta_j$:
\[
\hat\beta_{\text{shift-share}}=\sum_{j=1}^{m}\hat\alpha_j\hat\beta_j
\]
where the constants $\hat\alpha_j$, known as Rotemberg weights, depend only on the covariates and sum to $1$. $\hat\beta_j$ is not consistent if the $j$th share violates exogeneity. This implies that the shift-share estimate will be reasonably accurate as long as the shares with the largest absolute Rotemberg weights are exogenous. Researchers must be ready to defend the exogeneity of these shares more than others. R package \texttt{bartik.weight} and Stata package \texttt{bartik-weight} are available for weight calculation.

\paragraph{Panel settings} We write the full panel model as follows:
\[
Y_{it}=\beta X_{it}+\gamma^\top\Pi_{it}+\epsilon_{it},
\]
\[
Z_{it}=\sum_{j=1}^{m} w_{ij0}D_{jt}
\]
where $t=0,\ldots,T$ denotes time, $Z_{it}$ instruments $X_{it}$, and $\{\epsilon_{i0},\ldots,\epsilon_{iT}\}$ are independent across $i$.

Two points are worth noting. First, the model allows for temporal correlations in the error term but not spatial correlations across units, meaning it cannot accommodate geographically clustered errors. This is due to a lack of a theory of inference under clustering and overidentification (footnote 14 in \textcite{GoldsmithPinkhamSorkinSwift2020}), although \textcite{BorusyakHullJaravel2025b} recommend conventional clustering for practical purposes. Second, the instrument $Z_{it}$ fixes shares at their initial values. This is to avoid post-treatment bias that can arise if current shares were partly shaped by past shocks. \textcite{FoukaTabellini2022} use the initial spatial distribution of Mexican immigrants for this reason.

\paragraph{Diagnostic tests}Share exogeneity assumptions can be indirectly tested. First, the conditional covariance $\cov(w_{ij},T_i \mid \Pi_i)$ must be zero for shares $j$, especially with the largest Rotemberg weights, if $T_i$ proxies the error term $\epsilon_i$. $T_i$ can be any variables that affect the dependent variable not through the share instruments such as any labor supply shocks in \textcite{Bartik1991}. This can be practically done by regressing shares on covariates and interpret the regression results \parencite{GoldsmithPinkhamSorkinSwift2020}. Second, the estimate $\hat\beta$ should be zero in pre-treated periods if exist, analogous to pre-trend tests in difference-in-differences designs. The immigration regime change of $1965$ in \textcite{FoukaTabellini2022} is an example of the onset period. Third, since the shift-share instrument is one specific way to combine individual share instruments, the estimate should be robust to alternative ways to exploit multiple instruments. Overidentification tests formalize this idea \parencite{GoldsmithPinkhamSorkinSwift2020}.

\paragraph{Effect heterogeneity} The discussion so far has assumed units are homogeneous. How does the method fare under effect heterogeneity? Suppose the second-stage regression is replaced by
$Y_i=\alpha+\beta_i X_i+\epsilon_i$ 
where $\beta_i$ denotes heterogeneous treatment effects. It turns out that the shift-share estimate converges in probability to a weighted average of individual treatment effects $\beta_i$, but the weights are positive only if all unit-level effects are either uniformly positive or negative and all Rotemberg weights are positive.\footnote{If heterogeneity is also introduced in the first stage by writing 
$X_i=\sum_j w_{ij} t_{ij}D_j$ where $t_{ij}$ captures the true marginal effect of the $j$-th shift on the $i$-th unit’s outcome, the weights are convex only when shares are uncorrelated with one another. This condition is not attainable when shares are complete \parencite{HahnKuersteinerSantosWilligrod2024}.}
Since achieving both conditions is challenging, the shift-share regression may not provide a reliable estimate under effect heterogeneity.

\subsection{Shift Exogeneity: Many Exogenous and Independent Shifts}

Shift exogeneity holds when the shift distribution is mean-independent of the error term and shares. This framework applies when there are no viable selection-on-observables strategies for shares. It instead identifies conditions where comparable shifts can be combined into a single ``proper'' treatment. We treat shifts as random and independent, and everything else as fixed. A non-random sequence of shares and errors (a triangular array) will be considered in asymptotics. Formal discussions are relegated to Appendix.

\paragraph{Invalid instruments} Under shift exogeneity, the shift-share regression is the same IV regression as in share exogeneity but with \textit{invalid instruments} \parencite{KolesarChettyFriedmanGlaeserImbens2015}. Since shares are endogenous, the Rotemberg decomposition implies that the shift-share estimate is a weighted average of biased estimates. \textcite{AdaoKolesarMorales2019} conditions under which these biases cancel out analogously to the law of large numbers.

Two conditions are needed. First, shifts must be demeaned. This is to purge out the systematic variation in the shift-share variable. Consider the two-shift model from the previous section.

\begin{equation*}
\begin{aligned}
X_i
&= w_{i1}D_1 + w_{i2}D_2 \\
&= \underbrace{w_{i1}\mathbb{E}[D_1] + w_{i2}\mathbb{E}[D_2]}_{\text{systematic variation}}
+ \underbrace{w_{i1}(D_1-\mathbb{E}[D_1]) + w_{i2}(D_2-\mathbb{E}[D_2])}_{\text{stochastic variation}} .
\end{aligned}
\end{equation*}

As we are treating shares as fixed, units select into different aggregate treatments by the potentially endogenous share distribution unless $\mathbb{E}[D_1]=\mathbb{E}[D_2]=0$.\footnote{As noted in footnote 5, these expectations have already been conditioned on shares and errors by considering them fixed.}

Second, no shares can asymptotically dominate the others. This is for each endogenous share to contribute only a limited bias to the shift-share estimate so that the biases collectively cancel out. Shifts cannot also be strongly correlated as otherwise cancellation would fail from correlations among biases. Note that identification requires a large number of shifts as well as units unlike the share exogeneity framework.

\paragraph{Regression inversion} \textcite{BorusyakHullJaravel2022}, Hull and Jaravel (2022) propose an alternative way to understand identification. Consider the two-shift model again, with complete shares: $\sum_j w_{ij}=1$ for all $i$. Incomplete shares are discussed below. We first invert the regression turning shifts into ``observations.'' Taking a weighted average of the original second-stages $Y_i=\alpha+\beta X_i+\epsilon_i$ where each unit $i$ is weighted by its $j$-th share $\frac{w_{ij}}{\sum_i w_{ij}}$ yields

\[
\left\{
\begin{aligned}
\frac{\sum_i w_{i1}Y_i}{\sum_i w_{i1}} &= \alpha + \beta\frac{\sum_i w_{i1}X_i}{\sum_i w_{i1}} + \epsilon'_1 ,\\
\frac{\sum_i w_{i2}Y_i}{\sum_i w_{i2}} &= \alpha + \beta\frac{\sum_i w_{i2}X_i}{\sum_i w_{i2}} + \epsilon'_2 .
\end{aligned}
\right.
\]

This is a regression model with two ``observations.'' Note that the inversion does not affect the coefficients $\alpha$ and $\beta$. It turns out that the shift-share estimate is mechanically equivalent to the IV estimate in this inverted regression if shifts $D_1$ and $D_2$ instrument ``endogenous variables'' $\frac{\sum_i w_{i1}X_i}{\sum_i w_{i1}}$ and $\frac{\sum_i w_{i2}X_i}{\sum_i w_{i2}}$ with particular weights $\sum_i w_{i1}$ and $\sum_i w_{i2}$.\footnote{Weights here apply to units similarly to some geographical regressions that adjust for regional population size. Compare these with the GMM weights applied to overidentified moment conditions in the share exogeneity framework.}

Identification is achieved as long as shifts $D_j$ are orthogonal to the inverted errors $\epsilon'_j$ that consist of errors $\epsilon_i$ and shares $w_{ij}$. Since errors and shares are fixed, shifts must be not only mean-independent but also mean-zero to achieve the zero-correlation condition $\cov(D_j,\epsilon'_j)=0$. Consistency of the IV estimate further requires many uncorrelated shifts and restrictions on shares so that each regression weight $\sum_i w_{ij}$ becomes asymptotically negligible.

\paragraph{Inference} The equivalence result between the shift-share regression and the inverted regression applies only to the point estimate and not to the standard error. The typical heteroskedasticity-robust estimator in the inverted regression produces valid standard errors by the Central Limit Theorem if shifts are independent, or the cluster-robust estimator if shifts are clustered. Note that the inference remains valid under any dependence structure among shares and errors, and only the dependence structure of shifts matters in the inference procedure provided that shares satisfy negligibility. Available in R/Stata package \texttt{ssaggregate} and R package \texttt{ShiftShareSE}.

\paragraph{Control variables} If shifts are exogenous but centered around different means, one may directly model the means. Assume shifts are linear in controls such as shift-level covariates or dummies that represent ex-ante known shift groups:
\[
\mathbb{E}[D_j]=\gamma^\top p_j .
\]
The demeaned shift-share measure
\[
\sum_j w_{ij}(D_j-\gamma^\top p_j)=X_i-\gamma^\top\Bigl(\sum_j w_{ij}p_j\Bigr)
\]
would achieve identification if other assumptions are satisfied. $\gamma$ will be consistently estimated in the regression of shifts $D_j$ on their covariates $p_j$ as the number of shifts increases. Finally, estimating $\hat\gamma$ and demeaning $D_j$ is equivalent to controlling for an additional vector $\sum_j w_{ij}p_j$ in the shift-share regression by the Frisch--Waugh--Lovell theorem.

The inverted regression method also requires controls $\sum_j w_{ij}p_j$ before inversion, which are inverted using the same weights $\frac{w_{ij}}{\sum_j w_{ij}}$ as the shift-share measure. As a special case, this control scheme accommodates incomplete shares: $\sum_j w_{ij}<1$. Introduce a hypothetical shift $D_{m+1}$ that is identically zero with a complementary share $w_{i(m+1)}=1-\sum_j w_{ij}$. Define a shift-level dummy by setting $p_j=1$ for $1\le j\le m$ and $p_{m+1}=0$. Shares are now complete with the new shift, and the regression can be inverted with an additional control $\sum_{j=1}^{m+1} w_{ij}p_j=\sum_{j=1}^{m} w_{ij}$. This control is also needed in the invalid instrument approach.

\paragraph{Panel settings} We write the full panel model as follows:
\[
Y_{it}=\beta X_{it}+\gamma^\top\Pi_{it}+\epsilon_{it},
\]
\[
Z_{it}=\sum_{j=1}^{m} w_{ijt}D_{jt},
\]
where $t=1,\ldots,T$ denotes time and $Z_{it}$ instruments $X_{it}$. Rewrite the shift-share measure $Z_{it}$ in long form by including all $T\times m$ shifts across periods and assigning zero weights to shifts outside period $t$. Both approaches immediately apply without having to fix shares at their initial values. Clustering may be required if shifts are temporally correlated.

\paragraph{Diagnostic tests} Share conditions and shift correlation can be directly tested, while shift exogeneity assumptions can be indirectly tested similarly to share exogeneity. First,
\[
\cov\!\left(D_j,\frac{\sum_i w_{ij}T_i}{\sum_i w_{ij}}\right)=\cov(D_j,T'_j)=0
\]
for unit-level covariates $T_i$ and shift-level covariates $T'_j$ that proxy error terms $\epsilon_i$ and $\epsilon'_j$, after partialling out shift-level covariates $p_j$. Second, pre-trend tests if there exists the onset period. Third, \textcite{HahnKuersteinerSantosWilligrod2024} propose an overidentification test. The intuition is that if shifts are mean-independent of shares and errors, any functions of them cannot be correlated with the demeaned shifts, hence coefficients $\gamma$ to the shift-level controls $p_j$ being overidentified. Section~\ref{sec:china} illustrates diagnostic tests with an example.

\paragraph{Effect heterogeneity} The shift-share estimate $\hat\beta$ always converges in probability to a convex average of individual treatment effects when both the second-stage effects and the shifts are allowed to be heterogeneous. However, if randomness is introduced into shares as well, negative weights may occur unless every pair of shifts is positively correlated \parencite{HahnKuersteinerSantosWilligrod2024}.

\subsection{Shift Exogeneity: Finite Exogenous and Exchangeable Shifts}

The previous identification strategy had two major challenges. First, individual biases had to cancel out, which required a large number of shifts and a complex asymptotic analysis to justify consistency under arbitrary share and error distributions. Second, if shifts were not mean-zero, they had to be demeaned by modeling and estimating their means.

Both challenges can be bypassed by instead assuming that shifts are independent of errors conditional on shares and can be grouped so that they are exchangeable within each group (Borusyak and Hull 2023).\footnote{Exchangeability means that the joint distribution of shifts is invariant under permutation: $(D_1,\ldots,D_n)\overset{d}{=}(D_{\sigma(1)},\ldots,D_{\sigma(n)})$ for any permutation $\sigma$. This condition implies identical distributions but relaxes independence. For example, jointly normal shifts with a common correlation are not i.i.d.\ but exchangeable.}

The testing procedure consists of two steps. First, fix the effect size $\beta$ at some value. Second, randomize shifts and calculate the cross-moment between the shift-share instruments $Z_i$ and the residuals $Y_i-\beta X_i$ for the test statistic. Repeat the process for many values of $\beta$. The point estimate is the $\beta$ such that the test statistic under real shifts is the mean of the test statistics under randomization, and the confidence interval collects all $\beta$ such that the test statistic under real shifts is not too extreme in the test statistic distribution under randomization.\footnote{This randomization test is valid in more general settings where both instrument formula and the design are known, meaning that instruments are known functions of exogenous and endogenous components and the shift assignment process is prespecified.}

Asymptotics can be dispensed with since randomization inference is exact. Demeaning can be also dispensed with since it merely shifts each test statistic by a constant:
\[
\mathbb{E}\big[(\tilde Z_i-\mu_i)(Y_i-\beta X_i)\big]
=\mathbb{E}\big[\tilde Z_i(Y_i-\beta X_i)\big]
-\mathbb{E}\big[\mu_i(Y_i-\beta X_i)\big],
\]
where $\tilde Z_i$ is the recalculated instrument with randomized shifts and $\mu_i=\mathbb{E}[Z_i]$. Since the second term does not vary under shift randomization, demeaning does not affect the point estimation nor the interval estimation.

We conclude the section with three remarks. First, if the second-stage includes covariates, residualize the outcome variable $Y_i$ and the independent variable $X_i$ over the covariates before performing the permutation test (Appendix C.6). Second, although the test is exact under finite shifts, it imposes a stronger requirement that shifts be identically distributed, implying that all moments must coincide instead of the first moment in the previous section. Finally, as randomization inference tests the sharp null, results are harder to interpret under potential effect heterogeneity.

\subsection{Discussion}

Share exogeneity and shift exogeneity rely on different sets of identifying assumptions. The former exploits comparability among units, while the latter leverages comparability among shifts. Empirical designs that researchers have in mind may suggest which framework is more suitable. Share exogeneity is implied when they focus on similarity among units (through share exogeneity), or shocks to specific industries that are key to identification (those with potentially large Rotemberg weights). Shift exogeneity is relevant when they emphasize many comparable shocks or finite identical shocks unrelated to unit characteristics.

Share exogeneity might not be applicable if shares are codetermined with the outcome variable in a sort of equilibrium as in \textcite{AutorDornHanson2013} where both shares and errors have similar shift-share structures. The theory also does not yet accommodate geographically clustered errors. Shift exogeneity requires covariates that fully capture variation existing in shift means, as well as stronger assumptions on shifts such as mean-independence or identical distribution, in contrast to the zero correlation condition under share exogeneity. Neither framework provides fully satisfactory causal interpretation under effect heterogeneity.

What does one do if both shifts and shares are endogenous? One may consider exogenize one of them by fixing shares at their initial values \parencite{FoukaTabellini2022} or finding counterfactual exogenous shifts \parencite{AutorDornHansonMajlesi2020}. \textcite{HahnKuersteinerSantosWilligrod2024} explore the possibility that identification comes partly from shifts and partly from shares. In rare examples where both are exogenous, share exogeneity typically yields smaller standard errors than shift exogeneity as noted by \textcite{AdaoKolesarMorales2019}, and produces the same standard error as the shift-share regression by the equivalence result in Section~\ref{sec:4_1}.

\section{Shift-share designs in political science}
\label{sec:ps}
This section surveys the application of shift-share designs in political science, identifies the strengths and weaknesses of current studies, proposes a framework for understanding and developing shift-share designs, and discusses opportunities for future research. The review is based on Table A.1 that compiles thirty-five articles that use shift-share designs published in political science journals. These shift-share articles were identified through keyword searches for ``shift-share'' and ``Bartik'' on Google Scholar, as well as by tracking papers that cited milestone shift-share articles such as \textcite{Card2001} and \textcite{AutorDornHanson2013}. I note that creating an exhaustive list is challenging as some articles might have used shift-share designs as a measurement strategy without explicitly stating it.

\begin{table}[htbp]
\centering
\caption{Shift-Share Designs in Political Science Journals}
\label{tab:ps-shiftshare-journals}
\begin{tabular}{lccc}
\toprule
 & APSR/AJPS/JOP & Others & Total \\
\midrule
Before 2015     & 0 & 0  & 0  \\
2015--2019      & 4 & 2  & 6  \\
2020--present   & 8 & 21 & 29 \\
\bottomrule
\end{tabular}

\medskip
\begin{minipage}{0.9\linewidth}
\footnotesize
\emph{Note:} This non-exhaustive list was collected through keyword searches for ``shift-share'' and ``Bartik'' on Google Scholar and by tracking papers that cited milestone shift-share articles. needs update
\end{minipage}
\end{table}

Table~\ref{tab:ps-shiftshare-journals} presents a breakdown by year and journal. The first shift-share article was published in 2015, followed by five more before 2020. Twenty-nine articles have been published since then in the last five years including six forthcomings, indicating growing interest in shift-share designs within political science. The table also reveals another pattern: while shift-share designs were initially found in general-interest top journals, a growing number of articles are found in field-specific journals. This means that the shift-share methods are becoming a standard tool for studying specific topics.

The use of shift-share designs in political science largely mirrors their application in economics. Trade shocks are the most popular topic, with sixteen in total. Most adapt the research design of \textcite{AutorDornHanson2013} to alternative outcome variables or contexts; the notable exception is \textcite{BisbeeRosendorff2024} who create new trade shock measures using occupation-level data. Studies on technology shocks apply shift-share designs both for measurement and as instruments similar to trade shocks, but replacing industry-level import exposures with technological innovations \parencite{AutorDornHanson2015}. Immigration and migration follow with six studies, often adopting the Bartik instrument of \textcite{Card2001}. Capital movement, foreign aid and natural resources articles exploit the ``propensity score'' design introduced by \textcite{NunnQian2014}.

Despite the growing popularity of the methods, the associated identification problems are not as widely recognized in the field. Only one-third of the articles published in or after 2021 acknowledged potential en- dogeneity in shifts or shares and cited relevant source papers from the previous section. Only a few among those explicitly discussed identifying assumptions and conducted diagnostic tests, while most stopped at merely noting the possible methodological concerns. Moreover, the vast majority of studies relied on share exogeneity, knowingly or not, including articles on trade shocks that have been more associated with the shift exogeneity framework in economics. While this does not necessarily invalidate the designs, but the heavy reliance on share exogeneity suggests either a misapplication of the framework or an underutiliza- tion of shift exogeneity. These observations underscore the need for a more principled way to understand shift-share designs, particularly those less familiar with them.

\begin{table}[htbp]
\centering
\caption{Three Dimensions of Shift-Share Designs}
\label{tab:three-dimensions}
\begin{tabular}{lccc}
\toprule
Type & Measurement & Bartik & Propensity Score \\
\midrule
Components & Units & Shifts & Shares \\
Exogeneity & Share & Shift & Both \\
\bottomrule
\end{tabular}
\end{table}

Table~\ref{tab:three-dimensions} shows three basic dimensions to consider. The first dimension is the type of shift-share variables indicating why shift-share variables are needed in the first place. An example of measurement is the China shock where regional import exposure is approximated with regional employment shares and national import changes. This application of shift-share designs needs justification of the linear approximation to the unobserved true independent variable as well as the exogeneity of shifts or shares. Section~\ref{sec:examples} provides examples of the other two applications. Although the type does not directly affect the validity of research designs, it may help identify a shift-share variable relevant to the research as a first step.

The second dimension concerns the definition of the shift-share variable. While units are determined by the research question, researchers can be creative in selecting shifts and shares based on data availability. Section 3 lists examples that illustrate diverse designs at the end. Note that when the shift-share variable is used as an instrument, shifts and shares do not necessarily have a natural shift and share interpretation. The propensity score approach in \textcite{NunnQian2014} is an example, with more examples in Appendix A of \textcite{BorusyakHullJaravel2025b}. This formal definition of shift-share designs broadens their potential use compared to the decomposition-based substantive definition.

When one has decided on an overall measurement or instrumenting approach, it is good practice to write down the final variable in shift-share form. \textcite{FoukaTabellini2022} construct a Bartik instrument using the initial regional immigrant share $w_i$ with the national inflow $D_t$, but scaled by the predicted populatioñ $\tilde P_{it}$ of region $i$ at time $t$. Therefore, the actual instrument is
\[
Z_{it}=\frac{w_i D_t}{\tilde P_{it}}
\]
instead of $Z_{it}=w_iD_t$ (footnote 4). Since the predicted populatioñ $\tilde P_{it}$ varies across units, the 2SLS approach is valid only if $\frac{w_i}{\tilde P_{it}}$ can be viewed as exogenous shares conditional on shifts $D_t$. Express $w_i$ as the ratio of the initial numbers of regional to national immigrants $\frac{I_{i0}}{I_0}$ and note that the initial number of national immigrants $I_0$ is common across all instruments. It then suffices to establish that the ratios of the initial number of immigrants to the predicted population size $\frac{I_{i0}}{\tilde P_{it}}$, instead of the initial immigrant share $w_i$, are exogenous ``shares'' conditional on shifts $D_t$. Diagnostic tests and regression specification may be misled in general if shifts and shares are mischaracterized.

The third dimension is the source of exogeneity and the corresponding identification strategies. Section~4 and checklists in \textcite{BorusyakHullJaravel2025b} apply. Note that although certain topics tend to favor one identification strategy over the other in the literature, the link is not deterministic. \textcite{ScheveSerlin2023} identify the effect of their trade shock measure using share exogeneity while acknowledging most studies on trade shocks rely on shift exogeneity. If exogeneity can be argued for both shifts and shares as in \textcite{CarreriDube2017}, the analogy to difference-in-differences designs, and thereby reliance on the share exogeneity framework, suffices as standard errors remain unchanged without shift exogeneity. The empirical case where exogeneity comes from the combination of shifts and shares, as explored theoretically in \textcite{HahnKuersteinerSantosWilligrod2024}, remains to be seen.

Formalism can help clarify identification with non-standard shift-share variables. Consider Ziaja (2020) who extends the propensity score approach to multiple donors in the context of foreign aid:
\[
Z_{it}=\sum_j p_{ij}D_{jt}
\]
for $p_{ij}$ the fraction of years in which recipient country $i$ was aided by donor $j$ and $D_{jt}$ the total for- eign aid by donor country $j$ in year $t$. Rewrite $p_{ij}$ as ``fixed'' shares $p_{ij0}$ to rewrite the instrument into
\[
Z_{it}=\sum_j p_{ij0}D_{jt}.
\]
This is the typical panel shift-share structure under share exogeneity, and identification follows from standard assumptions that $\cov(p_{ij0},\epsilon_{it})=0$ for every $j$. Now extend shares $p_{ij}$ to $p_{ij\,t,t'}$ with additional year indices $t,t'$, setting $p_{ij\,t,t'}=p_{ij}$ if $t=t'$ and $0$ otherwise. The instrument then turns into
\[
Z_{it}=\sum_{j,t'} p_{ij\,t,t'}D_{jt'},
\]
which is the typical long-form structure in panel settings under shift exogeneity. Identification follows from suitable identifying assumptions.

I conclude the review with two suggestions for future research. First, shift-share designs can be a useful tool when treatments or instruments are correlated. Network studies have sometimes used the designs unknowingly \parencite{BorusyakHullJaravel2025b}. Imagine an experimental design where random nodes are seeded on a network and their neighbors select into the treatment depending on the fraction of seeders among their neighbors. Instruments are not independently determined as they arise from the interaction between the common seeder status and the underlying network structure:
\[
Z_i=\sum_j w_{ij}D_j
\]
where the share $w_{ij}$ is the inverse of the number of neighbors of node $i$, and the shift $D_j=1$ if node $j$ is seeded and zero otherwise. 2SLS may underestimate the standard error if the network was endogenously formed. Although not listed in Table A.1, similar examples may exist in political science that went unnoticed when the correlation structure needed to be accounted for. The standard error can be computed either analytically or via simulation under a given correlation structure among $Z_i$. One may alternatively use the shift exogeneity framework provided that the network has a sufficiently diffuse structure.

Second, shift-share designs may hold greater potential in the American politics literature. Only five of thirty-five articles in Table A.1 study American politics. This may reflect that topics such as trade and immigration where shift-share designs are common are less frequently studied in the American context, or that richer data reduces the need for such proxy variables. Nevertheless, I argue that the granularity of data makes shift-share designs have wider applicability in the field. One could measure legislators’ exposure to donor-level shocks via the donor network or industry-level shocks via the lobbying network -- the sheer number of donors and industries may constitute many independent shifts required in the asymptotic shift exogeneity framework. Endogenous policymaking across regions could be instrumented using fractional exposure to nationwide shocks. Behavioral outcomes such as racial agenda or sentiments often refract through racial composition in the region documented in the census. These examples show potential utility of the methods in core topics of American politics.

\section{Example: China shock}
\label{sec:china}
Given the low awareness of shift exogeneity, this section illustrates the framework by replicating \textcite{ColantoneStanig2018b}. See \textcite{FoukaTabellini2022} or \textcite{ScheveSerlin2023} for examples of share exogeneity. The authors ask how imports from China led to the rise of nationalism and far-right parties in European countries, using an identical shift-share design to \textcite{AutorDornHanson2013} apart from outcome variables and the geographical context. In light of the replication exercise of \textcite{BorusyakHullJaravel2022}, I highlight two additional procedures required due to the different geographical context: shift transformation and shift residualization.

\subsection{Research Design}

The authors regress electoral outcomes of electoral districts on the shift-share trade measure using data
from 15 European countries spanning from 1988 to 2007. The electoral outcomes of interest are (1) the
vote share-weighted mean and median nationalism score and the nationalist autarchy score of parties, and
(2) vote shares of far-right parties. The independent variable is the interaction between local manufacturing
industry employment shares and two-year differences in Chinese imports scaled by total national employment of the industry:
\begin{equation}
X_{rt}
=
\sum_{j=1}^{m}
\frac{L_{rj(t-2)}}{L_{r(t-2)}} \times
\frac{\Delta \mathrm{Import}_{cjt}}{L_{cj(t-2)}}
\equiv
\sum_{j',t'} w_{r j' t,t'} D_{j't'} .
\tag{3}
\label{eq:cs-design}
\end{equation}
where $c$ indexes countries, $r$ regions, $j$ industries, $t$ years. The last expression rewrites the shift-share
variable in long form where $j'$ indexes country-industries and $t'$ years (see Section~\ref{sec:ps}). Local employment
$L_{rjt}$ is measured at the NUTS-2 level, and industries $j$ are classified at the two-letter NACE level, 14 in
total, compared to 20 two-digit manufacturing SIC codes.\footnote{NUTS is the administrative geographical unit and NACE is the industry code system developed by Eurostat. \textcite{AutorDornHanson2013} use the 4-digit Standard Industrial Classification (SIC), which \textcite{BorusyakHullJaravel2022} find are clustered at the 3-digit level.}
The instrument replaces $\Delta \mathrm{Import}_{cjt}$ in Equation~\eqref{eq:cs-design} with two-year differences in Chinese imports
into the United States, $\Delta \mathrm{Import}_{US,jt}$. The authors aim to isolate the effect of ``exogenous changes in supply conditions in China, rather than \dots\ domestic factors that could be correlated with electoral outcomes.''\footnote{The authors list two possible sources of shift endogeneity. First, politicians may strategically protect certain districts from Chinese imports, in which case electoral outcomes determine the local exposure to trade with China. Second, both the shifts and the dependent variable may be affected by idiosyncratic local shocks such as economic fluctuation or political performance of incumbents that are not attributable to China.}
This clearly implies identification based on shift exogeneity, provided that correlations between the two import changes are primarily driven by supply shocks within China \parencite[2129--2130]{AutorDornHanson2013}
.

\paragraph{Endogeneity concern.}
Regions with different industry structures are likely to have different political orientation due to unique demographic composition, economic interests and historical backgrounds. These confounders may cause regions with similar industry structures to electorally respond to import shocks differently from regions with dissimilar industry structure. This observation has two implications. First, shares are likely correlated not only with levels of the error term but also with its changes in ways not easily addressed by conditioning on observables, making less feasible the application of share exogeneity as advocated in \parencite[p.~2598]{GoldsmithPinkhamSorkinSwift2020}. Second, for valid inference, errors should be clustered as implied by the share structure as well as at the NUTS-2 region-year level, as in the original article. The shift exogeneity framework takes shares and errors as fixed and exploits the randomness in shifts only, allowing for arbitrary correlations among shares and errors.

\paragraph{Unique challenges.}
This shift-share design has three structural differences from \textcite{AutorDornHanson2013}. First, both shift-share variables are measured at a higher level than the outcome variable as NUTS-2 regions contain multiple electoral districts: if $Y_i$ denotes the electoral outcome of district $i$, region $r$ corresponds to multiple regions $i$. The authors use district as the unit of analysis and cluster the errors by region-year. I show in Appendix~B.2 that the choice of the unit of analysis does not matter under shift exogeneity since the inverted regression is unaffected when districts and regions are properly weighted by their population. I pick region as the unit of analysis to better illustrate the effective sample size.

Second, the instrument suffers from a divide-by-zero problem in its shifts as Chinese imports to the US are divided by the national industry employment of much smaller European countries, some of which even lack the industries under analysis. Compare this with \textcite{AutorDornHanson2013} who divide the Chinese import to a group of comparably sized European economies by US industry employments. This motivates the shift transformation in the next section that neglects shifts with small denominators.

Third, electoral outcomes are not as frequently measured as independent variables and instruments, virtually creating a missing data problem. Note that identification and inference under shift exogeneity partly comes from the ability to isolate the stochastic part in realized shift values, or equivalently, to center shifts by properly modeling their means. I propose a residualization scheme that uses annual trade data and then estimates the effect size by regressing outcomes on the residuals.

\paragraph{Data reconstruction.}
The replication is based on the reconstructed data covering the shorter period from 2001 to 2007. The replication files
of the authors only contain the aggregate shift-share variables and not the individual shift and share components.
I reconstruct the individual shifts and shares using unlicensed, publicly available data following the protocol in
Appendix~C. The reconstructed independent and instrumental variables have the correlation of 0.87 and 0.90 with the
aggregate shift-share variables used in the original article.\footnote{The differences mainly stem from share estimates. While the original variables use national employment statistics sourced from each country, the reconstructed ones use Eurostat employment statistics that are noisier and limited in coverage.}
Table~F.1 replicates their main results using the original variables, the original variables censored from 2001 to 2007,
and the imputed and predicted reconstructed variables.\footnote{Imputed variables predict missing employments using linear regression, and predicted variables replace all observed values with values predicted by linear regression.}
Censored and reconstructed estimates were close enough, especially the effects on average nationalism scores. The following
analyses use the nationalism score for the main dependent variable. Consequently, the main lesson of the replication exercise
will be the importance of using the correct statistical procedures rather than a reversal of the authors' substantive findings.

\subsection{Shift Transformation and Residualization}

Shifts must be processed first if they are not mean-zero and independent. \textcite{BorusyakHullJaravel2022} propose residualizing the scaled import changes over period fixed effects and clustering shifts at a level lower than NUTS-2. 

I assume the following structural model for shifts:
\begin{equation}
\frac{\Delta \mathrm{Import}_{US,jt}}{L_{cj(t-2)}}
=\nu_{cj} + v_t + \eta_{cjt},
\tag{4}
\label{eq:cs-shift-model}
\end{equation}
where $u_{cj}(\equiv u_{j'})$ is the country-industry fixed effect, $v_t$ is the time fixed effect, and $\eta_{cjt}$ is the
mean-zero incidental deviation in the shift to be properly clustered. The residualization here is two-way to account for
the baseline differences in the denominator across country-industry pairs. Since changes in the employment are slower
than changes in the trade flow, I argue that country-industry fixed effects can reasonably control for the baseline
differences. The incidental term $\eta_{cjt}$ is expected to be independent across industries according to the findings of
\textcite{BorusyakHullJaravel2022}, but not across countries due to the common term $\Delta \mathrm{Import}_{US,jt}$.

Both \textcite{AdaoKolesarMorales2019} and \textcite{BorusyakHullJaravel2022} implicitly estimate the incidental term
$\hat{\eta}_{cjt}$ by controlling for aggregate fixed-effects $\sum_{j',t'} w_{rj't,t'} u_{j'}$ and $\sum_{j',t'} w_{rj't,t'} v_t$ in the
shift-share regression. This approach estimates the incidental component of the shifts only using those from the country-years
when elections were held, and therefore yields less precise estimates than directly obtaining $\hat{\eta}_{cjt}$ from the annual
trade data. Appendix~D extends estimation and inference in the shift-share regression to with residualized shifts
$\hat{\eta}_{cjt}$. I note that this estimation may come at the cost of the easy implementation of the inverted regression, but
the cost might be offset by dispensing with the need to introduce aggregate fixed-effects as in our example.

In addition to the residualization, I propose shift replacement to ensure the identical mean assumption. Since individual
European economies are much smaller than the U.S., the shifts in this study tend to be more volatile than those in \textcite{AutorDornHanson2013}. Moreover, some countries barely have certain industries, causing the divide-by-zero problem in
the shifts. This problem did not arise in the other paper as their sizes of their numerator and the denominator were comparable.
I replace shifts with zero if the sum of the regional shares of a certain industry in the country is less than 0.03, or if
$w_{cjt}=\sum_{r\in c} w_{rjt}$ is less than 0.03 with zero. 24\% of the shifts were replaced, most of which is in the petroleum
and nuclear fuel industry. Appendix~E provides theoretical justifications. Statistically, the replacement scheme better protects
the identical expectation assumption from possible misspecification of shift-level covariates and the measurement error from
data reconstruction. Substantively, the replacement scheme limits the analysis to industries that align more closely with the
structural model underlying the shift-share measure. Note that the scheme is not to replace all industry shocks with a national
employment share of less than three percent. This scheme can be applied to the share in the independent variable as well.

The estimation procedure is as follows. First, \textbf{MISSING FROM THE ORIGINAL DRAFT}

\subsection{Empirical Evaluation of Assumptions}

Asymptotic inference under shift exogeneity requires the mean-independence and the linear expectation of shifts, independence across shift clusters, and asymptotically negligible cluster shares, where shifts and shares refer to those in the instrumental variable. Mean-independence holds if the factors driving the trade shock between China and the U.S.\ do not affect the domestic politics of European countries. The other three assumptions are not discussed in \textcite{ColantoneStanig2018b}. The identical expectation and cross-cluster independence assumptions will depend on the way shifts are manipulated.

\paragraph{Shift independence.}
To test the shift independence conditions, I include all shifts from the countries and time period of interest regardless of whether elections occurred in a given country and year.\footnote{This is to secure test power, but might be problematic if election timing is correlated with the shifts. However, even if election timing is a function of trade shocks in parliamentary countries, the assumptions would not be violated as long as the function is not time-varying and does not depend on the past shifts.} Table~\ref{tab:shift_summary} reports shift summary statistics. The odd columns use replaced shifts and the even columns use residualized replaced shifts.\footnote{Shifts were weighted by their aggregate share in the process of residualization.} Shifts are centered around zero before residualization, and they show a high variance even after residualization. The standard deviation of residualized shifts is around three times of that in \textcite{AutorDornHanson2013}, as expected from the relative size of economies of the U.S.\ and the European countries. Residualization retains around 24 percent of the variation.

The lower panel tests the dependence among shifts. Autocorrelations measure the correlation between shifts at year $t$ and $t-1$,
or year $t$ and $t-2$. Intra-class correlation coefficients (ICC) estimate the following unweighted random-effect models:
\begin{align*}
\text{Unresidualized:}\quad
D_{cjt} &= \alpha + \beta_{ct}\times \mathbb{I}_{ct} + \gamma_{jt}\times \mathbb{I}_{jt} + \varepsilon_{cjt},\\
\text{Residualized:}\quad
D_{cjt} &= \alpha + (\beta_{ct}\times \mathbb{I}_{ct} \ \text{or}\ \gamma_{jt}\times \mathbb{I}_{jt})
+ \delta_{cj}\times \mathbb{I}_{cj} + \nu_t\times \mathbb{I}_t + \varepsilon_{cjt},
\end{align*}
where $\mathbb{I}$ are indicators and all random terms are normal.\footnote{Only one of $\beta_{ct}$ and $\gamma_{jt}$ is treated random and the other is fixed in the unresidualized model due to the insufficient number of years. The residualized model includes only one of $\beta_{ct}$ and $\gamma_{jt}$ due to collinearity, and $\delta_{cj}$ and $\nu_t$ are fixed.}
Large coefficients imply dependence within the groups and suggest the need for clustering. The results align with our specification.
Residualized shifts are uncorrelated at least two years apart, and exhibit little ICCs across country-year but high ICCs across
industry-year.\footnote{The apparent statistical significance of country-year ICCs is due to the asymmetric confidence intervals.}
One-year autocorrelations are significant because $\Delta US\ \mathrm{Import}_{jt}$ and $\Delta US\ \mathrm{Import}_{j(t-1)}$ both contain the
import change from year $t-2$ to year $t-1$. Since only one pair of elections in the sample was held one year apart in the same
country, 2002 and 2003 in the Netherlands, I drop the 2002 election from the data.

\begin{table}[!htbp]
\centering
\caption{Shift Summary Statistics}
\label{tab:shift_summary}
\begin{tabular}{lcccc}
\toprule
& (1) & (2) & (3) & (4) \\
\midrule
& \multicolumn{2}{c}{Imputed US shifts} & \multicolumn{2}{c}{Predicted US shifts} \\
\cmidrule(lr){2-3}\cmidrule(lr){4-5}
Mean & 6.808 & 0 & 6.827 & 0 \\
Standard deviation & 141.881 & 66.99 & 141.248 & 64.971 \\
\addlinespace
\textit{Specification} & & & & \\
\quad Residualized & F & T & F & T \\
\quad Replaced & T & T & T & T \\
\quad SSE &  & 0.242 &  & 0.23 \\
\addlinespace
\textit{Autocorrelations} & & & & \\
\quad 1-year & 0.882 & 0.609 & 0.91 & 0.665 \\
& (0) & (0.019) & (0) & (0.005) \\
\quad 2-year & 0.768 & 0.153 & 0.815 & 0.224 \\
& (0) & (0.581) & (0) & (0.383) \\
\addlinespace
\textit{Intra-class correlations} & & & & \\
\quad Country & 0.117 & 0.035 & 0.131 & 0.054 \\
& (0.03) & (0.018) & (0.039) & (0.025) \\
\quad Industry & 0.394 & 0.278 & 0.403 & 0.284 \\
& (0.056) & (0.053) & (0.046) & (0.046) \\
\bottomrule
\end{tabular}

\vspace{0.5em}
\begin{flushleft}
\footnotesize
\emph{Note:} Parentheses indicate $p$-values for autocorrelations and bootstrapped standard errors for intra-class correlations (ICC)
defined as the length of the 95\% confidence interval divided by $1.96\times 2$. Mean and standard deviation are weighted by shift
shares $w_{cjt}$. Columns (2) and (4) use residuals of shifts over country-industry and year fixed effects. SSE is the residual
variance compared to raw shifts. Autocorrelations report correlations between shifts one or two years apart. ICC report adjusted
random effects on country-year or industry-year indicators. Shifters are replaced by zero if $w_{cjt}$ is smaller than 0.03.
\end{flushleft}
\end{table}

\begin{table}[!htbp]
\centering
\caption{Shift- and Unit-level Placebo Test}
\label{tab:placebo}
\begin{tabular}{lccc}
\toprule
Variable & Estimate & SE & Obs \\
\midrule
\textit{shift-level:} & & & \\
\quad Initial \% of national industry employment & $-0.03^{***}$ & (0.012) & 1370 \\
\addlinespace
\textit{Unit-level:} & & & \\
\quad Initial \% of foreign-born population & 0.008 & (0.015) & 321 \\
\quad Initial \% of high-skilled workers & 0.658 & (0.999) & 335 \\
\quad Initial \% of high-technology workers & 0.121 & (0.248) & 335 \\
\quad Initial \% of medium- or low-skilled workers & $-0.062$ & (0.469) & 335 \\
\quad Initial \% of medium- or low-technology workers & 1.181 & (1.459) & 335 \\
\quad Initial \% of workers in primary sectors & $-1.265$ & (1.334) & 335 \\
\quad Initial \% of service industry workers & 2.683 & (3.529) & 335 \\
\bottomrule
\end{tabular}

\vspace{0.5em}
\begin{flushleft}
\footnotesize
\emph{Note:} The upper panel tests if all shifts in Table~\ref{tab:shift_summary} predict pre-shock shift-level variations using OLS in
the inverted regression. The lower panel tests if shifts that correspond to elections in the sample predict pre-shock unit-level
variations, with shifts instrumenting regional shocks. Covariates for the shift-level test are industry-country and year fixed effects.
Covariates for unit-level tests are regional-level country-year fixed effects following the original paper, plus newly added aggregated
shift-level fixed effects. All independent variables are normalized to variance 1. Standard errors are clustered by industry-year pair.
Imputed values are used whenever necessary. $^{*}p<0.1$; $^{**}p<0.05$; $^{***}p<0.01$.
\end{flushleft}
\end{table}

Table~\ref{tab:placebo}: to show exogeneity wrt the outcome variable, political proxies require additional data collection at subnational level\dots

\paragraph{Shift exogeneity.}
Table~\ref{tab:placebo} tests the shift mean-independence. The upper panel tests if shifts predict pre-shock shift-level variations,
and the lower panel tests if shifts corresponding to the elections predict pre-shock unit-level variations. Economic placebos being tested
are initial national employment shares by industry and initial worker composition by region. Political placebos could also be considered,
but they must come from subnational variation due to the residualization at the country-industry level.\footnote{Pre-shock political leaning or nationalism score would be some examples.}
These variables can affect the election outcomes independently of the trade shock and proxy the error term in the inverted regression that
the shifts are intended to instrument. All specifications use residualized shifts, or equivalently control for industry-country and year fixed effects.
Results do not reject the null hypotheses that instruments are uncorrelated with pre-shock variables, except in the case of the shift-level test.
The strong significance in the shift-level test might be an artefact of data availability. The true pre-shock period should be before 1988 when
China had hardly entered international trade, but the placebo variable relies on 1999 employment shares.\footnote{The mean and standard deviation of the initial national industry share are 1.09\% and 0.84\%, so the effect size can be substantially significant as well.}
Unit-level placebo tests use pre-1988 variables. The shift-level test is presented for illustrative purposes and should not be substantively interpreted
much. Table~F.2 reports similar results with unreplaced shifts.

\begin{table}[!htbp]
\centering
\caption {Cluster Share Distribution}
\label{tab:cluster_share}
\begin{tabular}{ccc}
\toprule
Max (share) & Max (share squared) & 1/HHI \\
\midrule
0.03 & 0.07 & 58.18 \\
\bottomrule
\end{tabular}

\vspace{0.5em}
\begin{flushleft}
\footnotesize
\emph{Note:} Max (share) and max (share squared) are bounded between 0 and 1, and the HHI is bounded above by the number of clusters.
Shares matching replaced shifts are included, but the complementary share is excluded. All metrics use imputed shares.
\end{flushleft}
\end{table}

\paragraph{Share negligibility.}
Given the industry-year shift clusters, asymptotic negligibility holds if the aggregate European economy is not dominated by certain industries
or if elections have occurred frequently enough during the given period. To assess the share negligibility assumption, I focus on regional
employment shares that match with the election data and calculate cluster shares. The total number of clusters is 98. Define cluster shares
$w_{jt}=\sum_{c} w_{cjt}$. Table~\ref{tab:cluster_share} presents share summary statistics. The first and second columns measure the ratio of the
largest $w_{jt}$ (or $w_{jt}^2$) and the sum of $w_{jt}$ (or $w_{jt}^2$). These two metrics show as how negligible the largest cluster is and should
be as close to zero as possible \parencite{AdaoKolesarMorales2019}. The third column measures the inverse of the sum of each cluster's relative
size squared, or the inverse of their Herfindahl Index (HHI). This shows the effect sample size in the inverted regression, so it should be as
large as possible \parencite{BorusyakHullJaravel2022}. All metrics indicate that resulting estimates will be consistent and asymptotically valid.

\paragraph{Pre-trend test.}
Not available, especially due to the data limitation.

\subsection{Replicated Results}

Given the absence of strong evidence failing falsification tests, I replicate the main results using the new shift-share estimator. Regression
specifications differ from the original paper in two ways. First, election outcomes are averaged by NUTS-2 region for congruence between the unit
of analysis and the unit of shifts. Second, shift-level controls are added to satisfy the identical mean assumption.

Table~\ref{tab:main_results} summarizes the main results. The dependent variables are the median and weighted average of nationalism scores in each
electoral district. Estimates measure how much an extra unit of Chinese import drove parties towards nationalism. The cluster row uses the same 2SLS
estimators as in the original paper where standard errors are clustered by NUTS-2 region-year pair. The BHJ row uses inverted regression estimators
following \textcite{BorusyakHullJaravel2022}, with standard errors clustered by industry-year pair.

\begin{table}[!htbp]
\centering
\caption{Effects of Chinese Imports on Party Nationalism}
\label{tab:main_results}
\begin{tabular}{lcccccc}
\toprule
& (1) & (2) & (3) & (4) & (5) & (6) \\
\midrule
& \multicolumn{3}{c}{Median} & \multicolumn{3}{c}{Center of Gravity} \\
\cmidrule(lr){2-4}\cmidrule(lr){5-7}
Cluster & 0.765$^{***}$ & 0.646$^{**}$ & 3.636 & 0.311$^{***}$ & 0.317$^{***}$ & $-0.735$ \\
& (0.277) & (0.264) & (7.227) & (0.113) & (0.117) & (2.134) \\
Obs & 3006 & 295 & 295 & 3006 & 295 & 295 \\
F & 1528.07 & 174.28 & 1.82 & 1528.07 & 174.28 & 1.82 \\
\addlinespace
BHJ & -- & 0.646$^{***}$ & 3.636 & -- & 0.317$^{***}$ & $-0.735$ \\
& -- & (0.207) & (10.981) & -- & (0.098) & (2.544) \\
Obs & -- & 335 & 335 & -- & 335 & 335 \\
F & -- & 67.46 & 0.13 & -- & 67.46 & 0.13 \\
\addlinespace
Unit of Analysis & District & Region & Region & District & Region & Region \\
shift controls & F & F & T & F & F & T \\
Country-Year FE & T & T & T & T & T & T \\
\bottomrule
\end{tabular}

\vspace{0.5em}
\begin{flushleft}
\footnotesize
\emph{Note:} Cluster reports 2SLS estimates with standard errors clustered by NUTS-2 region-year pair. BHJ reports inverted regression
estimates with standard errors clustered by industry-year pair per Borusyak, Hull and Jaravel (2022). Column (1) and (4) estimate the
electoral district-level regression as in the original paper with 2001--2007 election data. Column (2) and (5) convert the election data
to the NUTS-2 region level by taking the simple average of electoral outcomes. Column (3) and (6) add shift-level controls aggregated to
the regional level. All specifications use transformed shifts and imputed shares. $^{*}p<0.1$; $^{**}p<0.05$; $^{***}p<0.01$.
\end{flushleft}
\end{table}

A comparison between columns (1) and (2) and between columns (4) and (5) suggests that the results are not much affected by aggregation to the
district level. However, BHJ estimator reports much smaller effects and larger standard errors. The difference in points estimates are due to
the effect heterogeneity. While the other estimator by \textcite{AdaoKolesarMorales2019} would preserve the point estimate, it is not applicable
in this example since the number of shifts is larger than the number of units. Larger standard errors reflect the endogeneity problem between shares
and electoral outcomes. This suggests that regions with the similar industrial composition have correlated electoral outcomes unaccounted by trade shocks,
and NUTS-2 region-year clusters fail to capture this correlation by treating them as independent units.

Columns (3) and (6) control for shift-level fixed effects aggregated to the unit level, in addition to the unit-level region-year fixed effects in the
original specification. Our residualization strategy specifies that shifts are only comparable when partialled out by country-industry and time fixed effects.
These controls reflect the initial share of each industry and the total manufacturing share in the region when aggregated.\footnote{The controls are not collinear by design. Each aggregated country-industry fixed effect contains initial shares of the industry in the country across all years, while each aggregated time fixed effect contains all total manufacturing shares in the year across all countries and industries.}
A comparison between columns (2) and (3) and between columns (5) and (6) indicates that the results are not robust to the addition of these new control
variables either.\footnote{Note that this comparison is not affected by measurement errors caused by share imputation.}
In the shift exogeneity framework, the shift-share estimate balances out the error terms in the inverted regression (B.4) under the assumption that shifts are
independent and have the identical mean. The inconsistent results with respect to residualization suggest that the original results might have been driven by a
group of large, correlated shifts. The shift-share variable does not capture the correct impact of shifts if different means are not adjusted for. Also, there is
no way to account for dependence among shifts in conventional regression methods. Raw shifts are expected to have different means across country due to varying
economic sizes, and would be highly correlated within industry as they share the same US imports as a component. Table~F.3 shows that the results are similar
when predicted shares or raw shifts are used instead of imputed shares or transformed shifts.

\subsection{Discussion}

Take care of your shifts!

Again, Consequently, the main lesson of the replication exercise will be the importance of using the correct statistical procedures rather than a reversal of
the authors' substantive findings.

This section finds that the replication data meets assumptions required in the shift exogeneity framework but the results are not robust either to the asymptotically correct standard error estimator or shift residualization. The only methodologically valid estimates are those in columns (3) and (6) and BHJ row, and they suggest that the original findings are not well supported from the causal perspective. This replication exercise makes two original contributions. First, the residualization and clustering strategy proposed here is more sophisticated compared to \textcite{AutorDornHanson2013} that has an almost identical shift-share design. This is due to the complexity of the data structure that has three-way correlations compared to two-way in the original China shock paper. Second, I propose a shift transformation scheme that trims noisy outlier shifts that do not much affect the shift-share estimate. Shifts in this example involve division by small numbers that are not precisely observed. The scheme protects shifts from misspecifications, hence from erroneously rejecting the null hypothesis regarding the identical mean and independence assumption.

\section{Conclusion}
\label{sec:conclusion}
Having started as an imputation scheme, shift-share designs are now understood more generally as designs that involve inner-product variables between endogenous and exogenous variables. This article reviews how shift-share designs have been used in economics and political science from a methodological standpoint. Shift-share variables can measure factual unobservables using shares and observable shifts, or counterfactual unobservables using initial shares and counterfactual shifts such as the average value of the sample units or values from external units. Shift-share variables based on the propensity score are a statistical construct to enhance the relevance of instruments to the independent variable, but share the same statistical properties as other shift-share variables. This article synthesizes those statistical properties with unified notation and framework. Share exogeneity assumes units are comparable and independent similarly, but not entirely identical, to units in difference-in-differences. Shift exogeneity assumes shifts are comparable and independent so that errors balance out as in the law of large numbers.

Political science articles are increasingly adopting shift-share designs, but they rarely discuss or are fully aware of the required identifying assumptions. This article exemplifies how to assess those assumptions and how the new methods can affect results with a political science example. In the replication exercise, this article further proposes new methodological techniques that help to verify strict identifying assumptions that are otherwise not possible. Findings were compromised by either newly suggested standard error estimators or shift residualization, each indicating different identification issues underlying the original design. This article finds that despite complexities in shift-share designs, they still have significant potential in political science, particularly in American Politics where they are underutilized.

%\paragraph{Data Availability Statement}

\clearpage
\printbibliography
\clearpage

\appendix
%\onehalfspacing
\doublespacing
\setcounter{page}{1}
\setcounter{table}{0}
\setcounter{figure}{0}
\setcounter{equation}{0}
\setcounter{footnote}{0}
\renewcommand\thetable{A\arabic{table}}
\renewcommand\thefigure{A\arabic{figure}}
\renewcommand{\thepage}{S-\arabic{page}}
\renewcommand{\theequation}{A\arabic{equation}}
\renewcommand{\thefootnote}{A\arabic{footnote}}

\begin{center}
\textbf{\Large Online Supplementary Materials} 
\end{center}

% ======================================================
\section*{A \quad Shift-share Articles in Political Science}
% ======================================================
\setcounter{footnote}{0}

\begin{longtable}{p{0.12\linewidth} p{0.12\linewidth} p{0.05\linewidth} p{0.05\linewidth} p{0.60\linewidth}}
\caption{List of shift-share articles in political science journals}\label{tab:appendix:ss-articles}\\
\toprule
Paper & Topic & Cited & Type & Shift-share variable \\
\midrule
\endfirsthead

\caption[]{List of shift-share articles in political science journals (continued)}\\
\toprule
Paper & Topic & Cited & Type & Shift-share variable \\
\midrule
\endhead

\midrule
\multicolumn{5}{r}{\footnotesize Continued on next page}\\
\endfoot

\bottomrule
\endlastfoot

\textcite{feigenbaum2015} & trade shock & N &  &
``Specifically, following Autor et al. (2013a), we define import exposure per worker as...'' \\

\textcite{colantone2018a} & trade shock & N &  &
``...Autor, Dorn, and Hanson (2013) derive an empirical measure of regional exposure to
the Chinese import shock from a supply perspective... We employ the same empirical
approach.'' \\

\textcite{colantone2018b} & trade shock & N &  &
``To this purpose, we build a region-specific indicator for the exposure to Chinese imports
following the methodology introduced by Autor, Dorn, and Hanson (2013).'' \\

\textcite{thewissen2019} & trade shock & N &  &
``For our measure of exposure to Chinese import competition, we... measure this as the
value of the total imported goods as a share of the value added for sector $i$ in country $j$ in
year $t$.'' \\

\textcite{ballardrosa2021} & trade shock & N &  &
``We then constructed measures of local labor market exposure to import competition
equal to the change in Chinese import exposure per worker in a TTWA with imports
weighted in the TTWA by its share of national employment in a given industry (Autor et
al., 2013).'' \\

\textcite{kim2021} & trade shock & N &  &
``By leveraging geographical variation in industry specialization and national-level
variation in Chinese imports in the industry, they capture the exogenous shock from
China to the local economy.'' \\

\textcite{milner2021} & trade shock & N &  &
``I follow Autor et al. (2013), Colantone and Stanig (2018b), and others in defining the globalization shocks as...'' \\

\textcite{ballardrosa2022} & trade shock & N &  &
``[W]e define local labor market shocks as the average change in Chinese import
penetration in the commuting zone’s industries, weighted by each industry’s share in the
commuting zones’s initial employment.'' \\

\textcite{ferrara2023} & trade shock & Y & share &
``[W]e replicate the methodology developed by Autor et al. (2013)... we construct an index of exposure to import competition by US commuting zone.'' \\

\textcite{hosek2022} & trade shock & N &  &
``To estimate the trade stakes of a district, we used imports and exports across
manufacturing and other commodity industries weighted by employment in each
industry at the district level.'' \\

\textcite{scheve2023} & trade shock & Y & share &
``Although our measure of imports per worker is computed according to a shift-share
formula, our identification strategy does not rely on the use of exogenous variation in the
form of exports from Germany to a third party.'' \\

\textcite{vallprat2023} & trade shock & N &  &
``To account for economic grievances, I measure constituencies’ exposure to the colonial
trade shock using an indicator similar to the shift-share instrument developed by Autor
et al. (2013).'' \\

\textcite{bisbee2024} & trade shock & N &  &
``The change in imports per worker in a state or an industry (the shift) is allocated to
workers differentially—with more to those workers with higher immobility. A greater
share of the shift is allocated to workers whose jobs are more distant to other jobs in task,
geography, or industry space.'' \\

\textcite{dur2024a} & trade shock & N &  &
``Furthermore, they aggregated the weighted competitiveness of all industries to obtain a
single value that expresses the overall trade competitiveness of this region.'' \\

\textcite{dur2024b} & trade shock & N &  &
``To measure subnational trade competitiveness, following the approach outlined in detail
in Huber, Stiller and Dür (2023), we first calculate a country’s comparative advantage at
the industry group level.'' \\

\textcite{meyerroseforthcoming} & trade shock & N &  &
``The intuitive idea behind this approach is that local labour markets are differentially
affected by the growth in imports from low-wage countries depending on their prior
industry specialization.'' \\

\textcite{scholl2024} & technology shock & N &  &
``...[I]dentification stems from a shift-share approach, where we use pre-sample-period local employment composition to estimate the exposure to new technologies in a
time-varying fashion.'' \\

\textcite{finseraasforthcoming} & technology shock & N &  &
``The idea is that industry growth would happen in those municipalities that already had
investments in the industry. This constructed growth of the industry can be used as an
instrument for actual growth.'' \\

\textcite{finseraas2020} & immigration & N & share &
``[W]e construct a predicted immigrant inflow by distributing all incoming immigrants to
the BaC industry as if the initial licensing share of each trade completely determines the
allocation of the incoming immigrants.'' \\

\textcite{fouka2022} & immigration & Y & share &
``The instrument assigns decadal immigration flows from Mexico between 1970 and 2010
to destinations within the US proportionally to the shares of Mexican immigrants who
had settled there in 1960, prior to the change in the immigration regime introduced in
1965.'' \\

\textcite{lim2023} & immigration & N &  &
``I construct the instrument for regional emigration rates in Poland by interacting the unemployment rates in the United Kingdom (the exogenous pull factor) and the past emigration rates of each region in Poland before the EU accession.'' \\

\textcite{dipoppaforthcoming} & immigration & Y & share &
``I instrument migration using a shift-share instrument and I exogenously predict the shift component by leveraging droughts in the south of Italy as push factor for migration to the north.'' \\

\textcite{smoldtforthcoming} & immigration & N &  &
``We substitute a Bartik (1991) shift-share instrument for ours. In particular, we hold
exposure to vote shares as constant either in a single pre-treatment year or for a portion
of the pre-treatment period.'' \\

\textcite{xuforthcoming} & immigration & Y & share &
``[D]rawing on the literature for estimating the effect of the Great Migration of Blacks to
northern cities in the United States, I develop a shift-share instrumental variable of
predicted migration of the rural poor to cities in Brazil and to neighborhoods in São
Paulo.'' \\

\textcite{stubbs2020} & capital movement & N & share &
``Specifically, our instrument is the interaction of the within-country average of the number of conditions across the period of interest with the year-on-year IMF’s budget
constraint.'' \\

\textcite{brannlund2022} & capital movement & Y & share &
``I define market-risk exposure as the total value of risky assets in district $i$ during the year 1999, divided by the value of total assets in the same district, multiplied by the
change in the VIX index in period $t$.'' \\

\textcite{gavin2023} & capital movement & Y &  &
``Capital flows could be endogenous... We calculate the country’s share of global net capital flows and exclude the country’s immediate neighbors... again scaled to country
GDP.'' \\

\textcite{kernforthcoming} & capital movement & Y & shift &
``Our shift-share instrument is the multiplicative interaction between the number of
countries under programs and the long-run probability of a country being under IMF
programs.'' \\

\textcite{raessforthcoming} & capital movement & N &  &
``We leverage the share of inward FDI from HICs in each DC in our sample in the year
prior to the start of our panel (i.e., 2000) interacted with the difference between the GDP
growth rate of each country and the average GDP growth rate in Europe.'' \\

\textcite{ahmed2016} & foreign aid & N &  &
``The instrument interacts the legislative fragmentation of the U.S. House of Representatives with the probability a country receives U.S. aid in any year.'' \\

\textcite{ziaja2020} & foreign aid & N &  &
``I follow recent suggestions to interact exogenous variables on the donor side with endogenous recipient properties in order to increase cross-sectional variation.'' \\

\textcite{baccini2021} & employment & Y & share &
``Since layoffs are not randomly assigned, we develop an instrumental variables strategy
using shift-share methodology (\textcite{bartik1991}) derived from national layoff shocks,
weighted by initial county-level employment.'' \\

\textcite{dehdari2022} & employment & Y & share &
``I supplement the OLS analysis with an instrumental variable (IV) approach using a
Bartik instrument that predicts the number of layoff notices by the national trends in
notices within each industry, and the sectoral composition in each election precinct.'' \\

\textcite{baccini2023} & austerity & N &  &
``The key coefficient of interest is $\beta$, which estimates the interaction term between the
two main independent variables. It reflects how the impact of national-level austerity
measures varies across districts with different degrees of economic vulnerability.'' \\

\textcite{carreri2017} & natural resources &  & share &
``We assess whether changes in the international oil price exert differential impacts
among municipalities that produce more oil. Our cross-sectional variation is oil
dependence, defined as the value of oil produced in per capita terms in 1993.'' \\

\end{longtable}

\section*{B \quad Technical Details of Section 4}

\subsection*{B.1 Share Exogeneity: Exogenous Shares}

\textcite{GoldsmithPinkhamSorkinSwift2020} consider the following panel setup where $X_{it}$ has a constant linear effect on the dependent variable $Y_{it}$, and the instrument variable $Z_{it}$ has endogenous shifts and exogenous shares:
\begin{equation*}
Y_{it}=\beta X_{it}+\gamma^{T}\Pi_{it}+\epsilon_{it}
\end{equation*}
\begin{equation}
Z_{it}=\sum_{j=1}^{m}w_{ij0}D_{jt}
\label{b.1}
\end{equation}where t denotes time period $t=1,2,\cdot\cdot\cdot,T$ and $\Pi_{it}$ is a vector of control variables. \footnote{The original paper builds on Bartik regression in which shocks $D_{ijt}$ are decomposed into common components $D_{jt}$ and idiosyncratic components $\tilde{D}_{ijt}$ (e.g. $E[\tilde{D}_{ijt}]=0$ so $D_{jt}=\tilde{D}_{\cdot jt}$ and $X_{it}=\sum_{j=1}^{m}w_{ijt}D_{ijt}$), but here we discuss general shift-share designs.} With shifts $D_{jt}$ being fixed, unit-specific variables are i.i.d. across units, but not across time within the same unit. Therefore, the model accounts for any correlation structure in shifts $D_{jt}$ and any temporal dependence in $\epsilon_{it}$ but no geographical dependence. Note that $Z_{it}$ uses pre-treated shares $w_{ij0}$ to ensure exogeneity as shifts might affect shares over time.\footnote{ \textcite{BorusyakHullJaravel2022} note that fixing shares at their initial values can potentially weaken the instrument strength as $T\rightarrow\infty$. } The authors' core observation is that instrumenting $X_{it}$ with $Z_{it}$ is algebraically equivalent to instrumenting $X_{it}$ with $w_{i10},\cdot\cdot\cdot,w_{im0}$ using a particular weights. \footnote{
Since the first-stage relationship between $X_{it}$ and initial shares $w_{ij0}$ will vary over time, we need $m\times T$ instruments (initial shares interacted with time periods) in total where T denotes the total number of periods.} This implies that the shift-share design is valid if shares meet the conditions typically required for instrument variables. Before moving on, I point out that post-treated shares are unlikely to be exogenous in the panel OLS regression. This is because if $X_{it}$ comprises time-varying shifts and shares, the shares $w_{ijt}$ would be affected by previous shifts $D_{j0},\cdot\cdot\cdot,D_{j(t-1)}$. However, in case where shares happen to be exogenous or the data contains a single period, setting $Z_{it}=X_{it}=\sum_{j=1}^{m}w_{ijt}D_{jt}$ and instrumenting $X_{it}$ with $Z_{it}$ is algebraically equivalent to running an OLS regression of $Y_{it}$ on $X_{it}$.

The equivalence result is illustrated in Section I of the original paper. I build intuition through a simple example. Suppose that there are two industries and one time period and the model has no control variables. Then, the first-stage equation is
\begin{equation*}
X_{i} = \delta D_{1} w_{i1} + \delta D_{2} w_{i2} + \eta_{i}
\end{equation*}

Instrumenting $X_{i}$ on $Z_{i}$ is equivalent to instrumenting $X_{i}$ on $w_{i1}$ and $w_{i2}$ if the ratio of coefficients of $w_{i1}$ and $w_{i2}$ is restricted to $D_{1}/D_{2}$. If we take shares for instruments, the moment condition is $\hat{E}[w_{i1}\epsilon_{i}]=\hat{E}[w_{i2}\epsilon_{i}]=0$, or
\begin{equation*}
\sum_{i=1}^{n}w_{i1}(Y_{i}-\hat{\beta}X_{i})=\sum_{i=1}^{n}w_{i2}(Y_{i}-\hat{\beta}X_{i})=0.
\end{equation*}

Since the number of equations is greater than the number of variables, we minimize the weighted average of moment functions. If shifts $D_{1}$ and $D_{2}$ are given as weights,
\begin{equation*}
argmin_{\beta}\left[D_{1}\sum_{i=1}^{n}w_{i1}(Y_{i}-\beta X_{i})+D_{2}\sum_{i=1}^{n}w_{i2}(Y_{i}-\beta X_{i})\right]^{2}
\end{equation*}
\begin{equation*}
= argmin\left[\sum_{i=1}^{n}(D_{1}w_{i1}+D_{2}w_{i2})(Y_{i}-\beta X_{i})\right]^{2}
\end{equation*}which is the moment condition when $Z_{i}$ instruments. This shows that choosing shifts for GMM weights can restrict the ratio of coefficients of share instruments and recover the shift-share instrument estimate, though those weights will be generally suboptimal. \footnote{The resulting estimate will exhibit a higher variance than when using optimal weights determined by the data. }

Consistency of the shift-share estimate follows if the share instruments are valid. Two conditions are required. First, they must have non-zero correlation with the independent variables $X_{it}$ conditional on controls $\Pi_{it}$ for each t. Second, they must be exogenous to the dependent variable, i.e. $\mathbb{E}[w_{ij0}\epsilon_{it}|\Pi_{it}]=0$ for all i, j, t whenever the shift $D_{j}$ is non-zero. This exclusion restriction is analogous to the random assignment (or parallel trends) in difference-in-difference designs where control units and treated units are identical except for the treatment status. Here, common shocks $D_{j}$ affect all units but units cannot select into their degree of exposure conditional on control variables. \footnote{Researchers might want to consider controlling for aggregate shares. For example, if $w_{ij0}$ denote initial 4-digit code industry shares, initial 1 or 2-digit code industry shares of the location are likely to predict both 4-digit code industry shares and the dependent variable.} This allows us to attribute any difference in outcomes to the effect of the varying degrees of exposure to shocks and not to anything else. Second-stage controls apply to units; no tension in share exogeneity since controls are unit-level.

One concern is that shares are often equilibrium outcomes in which the dependent variable is simultaneously determined as in the case of \textcite{AutorDornHanson2013}, so they would not be exogenous in many cases. The authors recommend using first differences in the outcome of interest instead of its levels to address the problem. According to Adão, Kolesár and Morales (2019), however, $E[w_{ij0}\epsilon_{it}|\Pi_{it}]$ might not be zero even when first differences are used. This shows that identifying assumptions always must be justified in light of theories. In response, the authors identify several study designs that implicitly use the share exogeneity framework. Researchers likely invoke share exogeneity when they highlight the similarity among units apart from their differential exposure to common shocks. Alternatively, they do so when the emphasis is not on the multiplicity of industries but on a two-industry example or shocks to specific industries. This is because identification under the assumption of shock exogeneity requires a large number of shifts.

The authors further propose diagnostic tests to assess the share exogeneity assumption. First, researchers can examine if variables thought to affect the dependent variable via $\epsilon_{i}$ in equation~\ref{b.1} also predict share instruments. These variables would be correlates of local supply shifts in the settings of \textcite{Bartik1991}. A significant association between the variables and shares conditional on controls indicates imbalance of shares between different groups of units, suggesting endogeneity between the shift-share variable and the dependent variable.

Second, researchers can test whether the data exhibit pre-trends if variations in shifts started affecting the dependent variable from a certain time period onward. The dependent variable in such cases should have little correlation with individual share instruments as well as the shift-share variable. Third, researchers can exploit multiple instruments. As the shift-share instrument is one specific way to combine individual share instruments, the estimate should not differ significantly when shares are individually instrumented using 2SLS or alternative methods if the model is well-specified. Overidentification tests are a formal way to do this, but researchers must be aware of various assumptions underlying the different alternative methods.

When there are multiple instruments, not all of them influence the final estimate equally. Let $\hat{\beta}_{j}$ be the estimate instrumented with only the jth shares. It is known that for some constants $\hat{\alpha}_{j}\in R$ that sum to 1,
\begin{equation*}
\hat{\beta}_{shift-share}=\sum_{j=1}^{m}\hat{\alpha}_{j}\hat{\beta}_{j}
\label{eq:appendix:rotemberg-decomp}
\end{equation*}

$\hat{\alpha}_{j}$ are called Rotemberg weights. This implies that the shift-share estimate is most influenced by shares with the largest absolute weights, and researchers must be ready to defend exgoeneity of these shares more than others. The authors suggest presenting diagnostic test results with respect to these key shares along with results with respect to the shift-share instrument. \footnote{R and Stata packages for Rotemberg weights can be found at \url{https://github.com/paulgp/bartik-weight}.}

Finally, the above decomposition also provides insight into performance of the shift-share estimate in the presence of effect heterogeneity. Suppose the true model is:
\begin{equation*}
Y_{it}=\beta_{i}X_{it}+\gamma^{\top}\Pi_{it}+\epsilon_{it}
\label{eq:appendix:heterogeneous-effects}
\end{equation*}

so that the independent variable has unit-specific effects $\beta_{i}$. We can show that $\hat{\beta}_{j}$ converges in probability to a convex combination of $\beta_{i}$ if the true effect of the jth share instrument $w_{ij0}$ on the independent variable $X_{it}$ is positive across i or negative across i. However, this does not guarantee that $\hat{\beta}_{shift-share}$ converges in probability to a convex combination of $\beta_{i}$ as well since Rotemberg weights can be negative. How can researchers tell that $\hat{\beta}_{shift-share}$ is a convex combination of $\beta_{i}$ when Rotemberg weights are not consistently positive? We cannot estimate individual weights on $\beta_{i}$ since weights in $\hat{\beta}_{j}$ are not estimable. Variation in $\hat{\beta}_{j}$ instead provides some suggestive evidence. $\hat{\beta}_{j}$ assign different weights on $\beta_{i}$, so little variation in $\hat{\beta}_{j}$ suggests similar weights across units and negative Rotemberg weights in this case do not cause the negative weight problem. If $\hat{\beta}_{j}$ vary a lot, however, they can check if the patterns accord with researchers' substantive knowledge and then probe how shares with large negative Rotemberg weights might create negative weights in the final estimate. Regarding diagnostic tests under effect heterogeneity, overidentification tests might fail since $\hat{\beta}_{j}$ can vary even when shares are exogenous. The other two tests remain valid.

\subsection*{B.2 \quad Shift Exogeneity: Many Exogenous and Independent shifts}

\textcite{AutorDornHanson2013} and \textcite{BorusyakHullJaravel2022} take the exact opposite approach by treating shifts $D_{j}$ random and everything else non-random conditional on shifts. The baseline models of the two papers differ in whether the shift-share variable appear as the independent variable or the instrumental variable, but this does not impact identifying assumptions much as pointed out in the previous section. We consider the following shift-share OLS regression.

\begin{equation*}
Y_{i}=\beta X_{i}+\gamma^{\top}\Pi_{i}+\epsilon_{i}
\end{equation*}
\begin{equation}
X_{i}=\sum_{j=1}^{m}w_{ij}D_{j}, \quad \sum_{j=1}^m w_{ij} = 1
\label{b.2}
\end{equation}

We create a fictitious share $w_{i(m+1)}=1-\sum_{j=1}^{m}w_{ij}$ and a shift $D_{m+1}=0$ for each i if the original sum is smaller than one. The goal is to study the asymptotic properties of $\hat{\beta}$ as the number of shifts m grows when $D_{j}$ is repeatedly sampled with $\Pi_{i}$, $\epsilon_{i}$, $w_{ij}$ held constant. Studying conditional limiting distributions allows not only shares to be endogenous to the dependent variable but units also to have geographical dependence.\footnote{Temporal dependence is covered too when extended to panel settings, but spillovers are not.} The authors find that the shift-share estimate is consistent if shocks are as-good-as-randomly assigned and small enough as the number of shocks grows, but the conventional robust and clustered standard error estimators can vastly underperform. \textcite{AdaoKolesarMorales2019} observe that shares $w_{ij}$ can introduce a dependence between the independent variable $X_{i}$ and the error term $\epsilon_{i}$ that is not captured by typical clusters (Section III). Under the settings of \textcite{AutorDornHanson2013}, this means that economic zones with similar industry portfolio will have correlated errors even if they belong to different states. Shares likely create dependence when the dependent and independent variables are jointly determined by shares, causing the error term to contain shares as well. Taking first differences of the dependent variable does not solve the problem since first differences of shift-share variables are still shift-share variables. Fixing $\epsilon_{i}$ allows us to work around the complex dependence structure that arises in shift-share equilibria.\footnote{The authors introduce two possibly correlated but distinct shares for labor demand and labor supply each, further complicating the dependence structure in error terms.}

\paragraph{Inverted regression} proof of equivalence of identifying condition.

\textcite{BorusyakHullJaravel2022} find a similar equivalence result as above under shift exogeneity. Ignoring control variables $\Pi_{i}$ for now, the regression model~\ref{b.2} is consistent if $\sum_{i=1}^{n}X_{i}\epsilon_{i}/n$ converges to zero in probability. Observe that
\begin{equation}
\frac{1}{n}\sum_{i=1}^{n} X_i \epsilon_i = \frac{1}{n} \sum_{i=1}^n \sum_{j=1}^m w_{ij} D_j \epsilon_i = \sum_{j=1}^m D_j \frac{1}{n} \sum_{i=1}^n w_{ij} \epsilon_i = \sum_{j=1}^m \frac{w_j}{n} D_j \frac{\sum_{i=1}^n w_{ij} \epsilon_i}{w_j}
\label{b.3}
\end{equation}

where $w_{j}=\sum_{i=1}^{n}w_{ij}$ SO $\sum_{j=1}^{m}w_{j}=n$. \footnote{\textcite{GoldsmithPinkhamSorkinSwift2020} view $\sum_{i,j} w_{ij} D_j \epsilon_i = 0$ as shares $w_{ij}$ being invalid instruments, each of which has a non-zero covariance with the error term but those covariances average out \parencite{KolesarChettyFriedmanGlaeserImbens2015}}.If we rearrange equation~\ref{b.2} into the shift level as
\begin{equation}
\frac{\sum_{i}w_{ij}Y_{i}}{w_{j}}
=\beta\frac{\sum_{i}w_{ij}X_{i}}{w_{j}}
+\frac{\sum_{i}w_{ij}\epsilon_{i}}{w_{j}}.
\label{b.4}
\end{equation}

the last sum in equation~\ref{b.3} is the moment condition of the regression model~\ref{b.4} instrumented with $D_{j}$ and weighted by $w_{j}/n$
\begin{equation*}
\sum_{j=1}^{m}\frac{w_{j}}{n} \times D_{j} \times \frac{\sum_{i=1}^{n}w_{ij}\epsilon_{i}}{w_{j}} \xrightarrow{p} 0.
\end{equation*}

When does the above moment condition hold? The authors provide sufficient conditions. First, shifts are mean-independent and have the same mean, or $E[D_{j}]=\mu$ for all j conditional on other fixed parameters. Mean-independence means that shifts are exogenous to the model in practice, not necessarily identically distributed, but the same mean might be too strict a condition. This can be relaxed by introducing control variables as will be discussed below. Mean-independence implies
\begin{equation*}
E\left[\sum_{j=1}^{m}\frac{w_{j}}{n}D_{j}\frac{\sum_{i=1}^{n}w_{ij}\epsilon_{i}}{w_{j}}\right]=E[D_{j}]\times\sum_{j=1}^{m}\frac{w_{j}}{n}\frac{\sum_{i=1}^{n}w_{ij}\epsilon_{i}}{w_{j}}=0.
\end{equation*}

Second, shifts are independent to one another and shares $w_{j}$ become negligible as their number grows.\footnote{Consistency of the shift-share estimate holds when shifts are uncorrelated or weakly correlated with one another, but the asymptotic requires independence (Assumption B5). I focus on inference rather than consistency as most political science works are interested in inference beyond identification.
} Intuitively, the variance of the moment condition is the sum of the variance of each term if shifts are independent, and each variance should be asymptotically negligible if so are weights:
\begin{equation*}
V\left[\sum_{j=1}^{m}\frac{w_{j}}{n}D_{j}\frac{\sum_{i=1}^{n}w_{ij}\epsilon_{i}}{w_{j}}\right]=\sum_{j=1}^{m}\left(\frac{w_{j}}{n}\frac{\sum_{i=1}^{n}w_{ij}\epsilon_{i}}{w_{j}}\right)^{2}V[D_{j}] \xrightarrow{p} 0.
\end{equation*}

This equivalence result shows that the shift-share point estimate equals the IV point estimate instrumented with shifts using particular weights. However, only the IV regression produces a valid standard error under the heteroskedasticity-robust estimator since unlike in the previous section, shifts no longer instrument the original regression model. \textcite{AdaoKolesarMorales2019} derive the consistency and asymptotic distribution of the OLS estimate under almost identical assumptions directly from equation ~\ref{b.2}. \footnote{$\max_j w_j^2 / \left(\sum_j w_j^2\right) \rightarrow 0$ is assumed in both papers, but these authors assume 
$\max_j w_j / \left(\sum_j w_j\right) \rightarrow 0$ 
instead of $\max_j w_j \rightarrow 0$. The fictitious share is excluded in these share assumptions.} Both approaches yield the same estimator if the model contains no control variables. 

Including controls under shift exogeneity is a bit complicated since mean-independence must hold at the shift level while many controls would be available at the unit level. We define a latent variable vector $\eta_{j}$ and a fixed vector $\mu$, and $\mathbb{E}[D_{j}|\eta_{j}]=\eta_{j}^{T}\mu$ for all j conditional on other fixed parameters. However, the two papers differ in how the latent variables $\eta_{j}$ are mapped to unit-level control variables. Borusyak, Hull and Jaravel (2022) assume that the control variable set $\Pi_{i}$ contain the exact average of the latent variables weighted by shares: if we denote the kth element in $\eta_{j}$ by $\eta_{jk}$ and the kth control by $\Pi_{ik}$, $\sum_{j=1}^{m}w_{ij}\eta_{jk}=\Pi_{ik}$ for all i. \textcite{AdaoKolesarMorales2019} allow asymptotically negligible noises in the controls: for small noises $u_{ik}$, $\sum_{j=1}^{m}w_{ij}\eta_{jk}+u_{ik}=\Pi_{ik}$ for all $i$. \footnote{$\sigma_k^2 / n \rightarrow 0$ and $\sigma_k^2 / \sqrt{\sum_j w_j^2} \rightarrow 0$ where $\sigma_k^2 = \sum_i \mathbb{E}[u_{ik}]^2$. The authors derive exact asymptotics under these assumptions, while \textcite{BorusyakHullJaravel2022}show that their standard errors are asymptotically conservative in the presence of any form of errors although consistency might not be guaranteed.
} The two papers reach slightly different asymptotic distributions due to a difference in estimating procedures, but both are asymptotically valid under the assumptions in each paper. \footnote{R package \texttt{ShiftShareSE} and R/\texttt{stata} package \texttt{ssaggregate} compute these standard errors. \texttt{ShiftShareSE} also provides null-imposed standard errors, which are known to perform better in finite samples, or in the IV regression, when the shift-share instrument is weak.} Their setups imply that the sum of real shares has to be controlled for when shares are incomplete. Let us denote the fictitious share and shift by index $m+1$ and define $\eta_{j}=1$ for $1\le j\le m$ and $\eta_{j}=0$ for $j=m+1$ Since $E[D_{1}],\cdot\cdot\cdot,E[D_{m}]$ would not equal $E[D_{m+1}]=0$ in general, $\eta_{j}$ distinguishes the fictitious shift from the real shares. The corresponding control $\sum_{j=1}^{m+1}w_{ij}\eta_{j}=\sum_{j=1}^{m}w_{ij}$ is the sum of real shares for each unit. Further heterogeneity existing among the real shifts can be controlled with additional latent variables that take a value of zero for the fictitious share.

\paragraph{Second-stage controls} shift-level controls vs unit-level controls in shift exogeneity. \textcite{ChetverikovHahnLiaoSantos2023} derive a simpler and asymptotically equivalent point and standard error estimators under the shift independence. 

Nevertheless, control variables might not be able to capture all dependence among shifts, especially in panel settings where shifts may be serially correlated. We can cluster shifts in such a case. Analogously to the random effect model, \textcite{AdaoKolesarMorales2019} derive asymptotics under the assumptions that shifts are independent only across different clusters and the largest cluster size is asymptotically negligible. Clusters can be defined over both time and shifts in their model. Meanwhile, \textcite{BorusyakHullJaravel2022} cluster shifts by imposing temporal or cross-sectional dependence on the error terms of the shift-level IV regression. \footnote{As a reminder, this procedure has nothing to do with the dependence structure in the error terms of the original OLS regression as the entire analysis has already been conditioned on all parameters but shifters.} Another notable issue that might arise in panel settings is fixed effects. Any relevant control variables in addition to those that control for latent variables, including two-way fixed effects, are innocuous to identification and increase efficiency if the model is correct. However, unit-fixed effects in the unit-level regression cannot remove time-invariant components in shifts unless shares are invariant across time.

\paragraph{SE estimators} Researchers can run similar diagnostic tests to assess the shift exogeneity assumption. Since units and shifts are at the different level, balance tests can be done at both the unit level and the shift level. Unit-level balance tests ask if variables that are thought to affect the dependent variable via $\epsilon_{i}$ in equation ~\ref{b.2} predict the shift-share variable $X_{j}$. shift-level balance tests ask if variables that are thought to affect the shift-level dependent variable via the error term in equation~\ref{b.4} predict shift instruments $D_{j}$, where shocks must be weighted by $w_{j}$.

Clustering can be applied likewise when shifts have remaining dependence after conditioned on latent variables. Pre-trend tests are also available if shifts started affecting the dependent variable from a certain time period onward. On the other hand, overidentification test is not possible unless the model contains two or more shift-share instruments as the equivalence result shows the same number of endogenous variables and instruments.

The authors lastly discuss three additional issues. First, consider data generated by the following heterogeneous effect model:
\begin{equation*}
Y_{i}=\sum_{j=1}^{m}\beta_{ij}w_{ij}D_{j}+\gamma^{\top}\Pi_{i}+\epsilon_{i}.
\end{equation*}
If we use the homogeneous effect model~\ref{b.2} and~\ref{b.4} for estimation, $\beta$ in each model converges to different convex combinations of $\beta_{ij}$. While standard errors will be conservative, these estimates might not be quantities of interest as neither is a convex combination of treatment effects, which are $w_{ij}\beta_{ij},$ not $\beta_{ij}$. Different estimates under the two estimators might suggest effect heterogeneity and hence a threat to inference.

Second, the above arguments hold when we consider shift-share IV regression instead of shift-share OLS regression. The equivalence result holds if we replace $X_{i}$ with the instrumental variable $Z_{i}$ in equation ~\ref{b.3} so that the shifts in the instrument variable are as-if random. If we do not use the equivalence result, the estimation procedures and inference parallel those of the regular IV estimator.

\paragraph{
Estimated shifts } in the shift-share IV settings, sometimes the shift $D_{j}$ has to be estimated instead of directly observed. For example, \textcite{bartik1991} assumes that the local employment share is the sum of the national labor demand shocks and the local labor supply shocks, and the national employment share proxies the national labor demand shocks since the local labor supply shocks would average out. This can be formally presented as
\begin{equation*}
D_{ij} = D_j + v_{ij}, \quad \hat{D}_{j}=\sum_{j=1}^{m}w_{ij}D_{ij}
\end{equation*}
where $D_{ij}$ is the observed local employment share that contains a national shock $D_{j}$ and local shocks $v_{ij}$. One concern is that since $D_{ij}$ is endogenous to local shocks, the instrument might be endogenous to the jth independent variable $X_{j}$ if it contains $D_{ij}$ in it. The authors find that consistency holds if the number of units is much larger than the number of shifts, but if it is not the case, the leave-one-out estimator ($\hat{D}_{j,-k}=\sum_{i\ne k}w_{ij}D_{ij}$ and the instrument $Z_{k}=\sum_{j}w_{kj}\hat{D}_{j,-k}$ for unit k) performs better. \footnote{More shifts may need to be dropped if $u_{ij}$ has geographical dependence.}

\paragraph{Overidentification tests} \textcite{HahnKuersteinerSantosWilligrod2024} Although inverted regression is ostensibly just-identified, we can convert the mean-independence to infinitely many zero correlation conditions by considering arbitrary functions shares and errors.

\paragraph{Unit of analysis} Section 6.1 discusses the "correct" unit of analysis in shift-share designs. Assume that the true trade exposure consists of unobservable district employment shares and national trade shocks: $X_{i}=\sum_{j}w_{ij}D_{j},Z_{i}=\sum_{j}w_{ij}D_{j}^{\prime}$. The outcome is generated through a simple regression: $Y_{i}=\alpha+\beta X_{i}+\epsilon_{i}$ where $E[\epsilon_{i}|\{w_{ij}\}, \{D_{j}\}] = 0$ under total exogeneity (which the original article arguably predicates on), and $E[D_{j}^{\prime}|\{w_{ij}\},\{\epsilon_{i}\}]=0$ under shift exogeneity. Each region has the same number of districts, and observable regional shares $w_{rj}$ are the simple average of all district shares $w_{ij}$ in the region. \footnote{One can weight districts by their population if districts and regions are of different sizes.} Trade exposure of district i is proxied with $X_{r}=\sum_{j}w_{rj}D_{j}$, $Z_{r}=\sum_{j}w_{rj}D_{j}^{\prime}$ for region r such that $i\in r$. 

If we also aggregate the outcome variable by region so that $Y_{r}=\alpha+\beta X_{r}+\epsilon_{r}$ then the identifying assumptions still hold under each exogeneity mode, i.e., $E[\epsilon_{r}|\{w_{rj}\}, \{D_{j}\}] = 0$ and $E[D_{j}^{\prime}|\{w_{rj}\}, \{\epsilon_{r}\}\}=0$, and $\beta$ is identified via the 2SLS/inverted regression of $Y_{r}$ on $X_{r}$ and $Z_{r}$ What if the district-level outcome $Y_{i}$ is regressed on regional proxies $X_{r}$ and $Z_{r}$?

The 2SLS under total exogeneity solves for the moment conditions $\hat{E}[Y_{i}-\alpha-\beta X_{r}]=\hat{E}[(Y_{i}-\alpha-\beta X_{r})Z_{r}]=0.$ Simple algebra shows $\hat{E}[Y_{i}]=\hat{E}[Y_{r}]$ and $\hat{E}[Y_{i}Z_{r}]=\hat{E}[Y_{r}Z_{r}]$, implying that identification under this 2SLS is equivalent to the regional-level 2SLS. Inference would be asymptotically correct with correctly clustered standard error estimates under total exogeneity as the number of regions grows to infinity. Under shift exogeneity, the inverted regression instruments with $D_{j}^{\prime}$ the inverted error $\frac{\sum_{i}w_{rj}Y_{i}}{\sum_{i}w_{rj}}-\alpha-\beta\frac{\sum_{i}w_{rj}X_{i}}{\sum_{i}w_{rj}}$ which is equivalent to the error inverted from the regional-level regression $\frac{\sum_{r}w_{rj}Y_{r}}{\sum_{i}w_{rj}}-\alpha-\beta\frac{\sum_{r}w_{rj}X_{r}}{\sum_{r}w_{rj}}$ Both identification and inference follow from the equivalence of the shift-level structural model rather than the equivalence of the identifying moment conditions.

\subsection*{B.3 Discussion}

Researchers likely invoke share exogeneity when they highlight the similarity among units apart from their differential exposure to common shocks. Alternatively, they do so when the emphasis is not on the multiplicity of industries but on a two-industry example or shocks to specific industries. This is because identification under the assumption of shock exogeneity requires a large number of shifts. The previous sections discussed how shift-share variables identify the underlying parameter when they consist of one endogenous variable and one exogenous variable. Identification strategies depend on whether shares are exogenous or shifts are exogenous.

When shares are exogenous, shift-share designs can be interpreted as difference-in-difference where otherwise identical units are treated with the same set of shifts but to exogenously determined degrees. shifts can be endogenous in the sense that their expectation may depend on the share or error term distribution among units. The 2SLS estimator can be shown to yield valid estimate and standard errors by converting the shift-share design into IV regression. The share exogeneity assumption need to be defended harder for some shares than others accordingly to their respective influence on the final estimate.

When shifts are exogenous, shift-share designs gather many negligibly small shocks and estimate the underlying parameter by averaging their effect. Shares may be endogenous as long as shocks are not seemingly affected by any other unit-level characteristics including shares or outcomes. The OLS or 2SLS estimator yield a valid point estimate if effects are homogeneous but tends to vastly underestimate standard errors. This is because shares create dependence among unit-level outcomes that are not accounted in typical clustering procedures.

The papers suggest two alternative estimators that work under slightly different assumptions, but they should not differ much under the condition where both sets of assumptions hold. \footnote{One important difference not mentioned above is that \textcite{AdaoKolesarMorales2019} require more units than shifts. Hahn et al.\ (2024) bypasses this problem by estimating the ridge regression rather than the OLS.} In addition to shift exogeneity, researchers must establish that there are an enough number of independent clusters of shifts so that their shares can be treated asymptotically negligible.

One concern is that shares are often equilibrium outcomes in which the dependent variable is simultaneously determined as in the case of \textcite{AutorDornHanson2013}, so they would not be exogenous in many cases. The authors recommend using first differences in the outcome of interest instead of its levels to address the problem. According to \textcite{AdaoKolesarMorales2019}, however, $E[w_{ij0}\epsilon_{it}|\Pi_{it}]$ might not be zero even when first differences are used. This shows that identifying assumptions always must be justified in light of theories.

In response, the authors identify several study designs that implicitly use the share exogeneity framework. Researchers likely invoke share exogeneity when they highlight the similarity among units apart from their differential exposure to common shocks. Alternatively, they do so when the emphasis is not on the multiplicity of industries but on a two-industry example or shocks to specific industries. This is because identification under the assumption of shock exogeneity requires a large number of shifts.

Both schemes come with restrictions in their models that researchers have to be mindful of. The first scheme ignores spatial correlation, a factor typically considered in panel analyses through clustering. \footnote{Since clustering under many instruments has not been studied, we do not know if clustered standard errors under many share instruments would be equal to clustered standard errors under a single shift-share instrument.} The second scheme introduces strict assumptions on a large number of shifts and non-standard assumptions regarding control variables. Neither of them performs effectively if shift-share variables exhibit heterogeneous effects. Also, shift-share designs inherit common problems in OLS or IV regression designs such as outliers and weak instruments.

What if both components of a shift-share variable are endogenous or exogenous? If both are endogenous, one may consider exogenize either of them by fixing shares at their initial values or averaging shifts as in \textcite{bartik1991}. If both are exogenous, the share exogeneity scheme suffices to justify the design. It remains an open question if there is a more efficient way to take advantage of exogeneity of both components, although this would not be the case in most applications of shift-share designs.

\newpage

\section*{C \quad Data Reconstruction in Section 6 }

\begin{table}[h!]
\caption*{Table C.1: Data source} 
\begin{tabular}{p{0.22\textwidth} p{0.26\textwidth} p{0.46\textwidth}}
\toprule
Variable & Source & Comment \\
\midrule

Trade Flow & Eurostat Comext & \\

Trade Flow (US, Norway) & UN Comtrade & Needs Comtrade API to download \\

Product Crosswalk & Eurostat Comext & The first two digits of CPA2002.txt were used for NACE Rev.\ 1.1 division. \\

CPI & OECD & Consumer price indices (CPIs, HICPs) (COICOP 1999) \\

Regional Employment & Eurostat SBS & SBS data by NUTS 2 region and NACE Rev.\ 1.1 (sbs\_r\_nuts03) \\

Total Regional Employment & Eurostat LFS & Employment by sex, age and NUTS 2 region (lfst\_r\_lfe2emp) \\

Election Outcomes & Replication data & Analysis\_Dataset\_District\_Level.dta \\

\bottomrule
\end{tabular}
\end{table}

\paragraph{Data compatibility}
About 4 percent of the HS-6 product codes in UN comtrade data belong to more than one NACE industries.
Volumes for these products were equally divided into each industry. Different data sources from Eurostat
contain different versions of the NUTS-2 system. Those regions were matched by their names in such
cases as most codes inherited the older names. UKI codes were refined into smaller regions over time.
I use the older, larger regions for analysis to minimize confusion. Variables for these regions were
simple averaged.

\paragraph{Imputation method}
Some regional industry employment statistics are missing in the data, which precludes the full
reconstruction of the replication data. I use the following linear regression for imputation:

\[
L_{rjt} = \alpha
+ \sum_c \beta_c \mathbb{I}_c
+ \sum_j \beta_j \mathbb{I}_j
+ \sum_{c,j} \beta_{cj} \mathbb{I}_{cj}
+ \sum_t \gamma_t \sigma_r \mathbb{I}_t
+ \sum_j \gamma_j \sigma_r \mathbb{I}_j
+ \sum_{t,j} \gamma_{tj} \sigma_r \mathbb{I}_{tj}
+ \varepsilon_{rjt}
\]where $\mathbb{I}$ denotes indicator and $\sigma_r$ is the standard deviation of all observed employments
in region $r$. Imputed shares use the imputation scheme only for missing values, while predicted shares
replace observed values with the imputed ones. Adjusted $R^2$ is 0.99. National employments by industry
are the sum of the regional employments:

\[
L_{cjt} = \sum_{r \in c} L_{rjt}.
\]

\newpage 

\section*{D \quad Shift Residualization}

This section develops a standard error estimator for the inverted regression that directly utilizes residualized shifts. 

We begin with the standard structural model under shift exogeneity as in Section 4.2.

\noindent \textbf{Assumption D.1.} $X_{i}=\sum_{j=1}^{m}w_{ij}\pi_{ij}D_{j}+v_{i}$ and $Y_{i}=\beta X_{i}+\gamma^{\top}p_{i}+\epsilon_{i}$

Note that no assumptions are imposed on the error terms $v_{i}$ and $\epsilon_{i}$ besides the regularity condition below. This is to ensure that $D_{j}$ is the only source of randomness that gives identification in the model. We implicitly assume that $X_{i}$ does not have to strictly follow the same shift-share structure as the instrument $Z_{i}=\sum_{j=1}^{m}w_{ij}D_{j}$, but it cannot be so far from the structure that $v_{i}$ violates the required regularity assumptions below.

\noindent \textbf{Assumption D.2.} For $j^{\prime}=1,\cdot\cdot\cdot,m^{\prime}(\ge m)$ and $\mathcal{F}_{m}=((\mu_{j^{\prime}})_{j^{\prime}},(\epsilon_{i},v_{i},p_{i},w_{ij},\pi_{ij})_{j})_{i}),$ shifts $D_{j^{\prime}}$ satisfy $D_{j^{\prime}}=f(\mu_{j^{\prime}})+\eta_{j^{\prime}}$ and $E[\eta_{j^{\prime}}|\mathcal{F}_{m}]=0$ for a fixed function $f:\mathbb{R}^{r}\rightarrow \mathbb{R}$.

Assumption D.2 extends the previous structural model in two ways. First, it allows more shifts to be observed than those actually used in the shift-share regression so that the incidental term can be better estimated. One motivation is that $D_{j}$ has a panel structure with two-way fixed effects and the shift-share regression includes only a subset of the years in which shifts are observed. Second, shifts $D_{j}$ may be nonlinear in the latent variables $\mu_{j}\in\mathbb{R}^{r}$ implying that researchers are free to use any estimation method provided that the estimator satisfies the required regularity assumptions below. $\mathcal{F}_{m}$ denotes parameters to be conditioned on along the triangular array of populations. To avoid confusion, index $j^{\prime}$ applies to all observed shifts, and $j$ to shifts included in the shift-share regression.

The following assumptions are inherited from Proposition 5 of \textcite{BorusyakHullJaravel2022}, except for Assumption B4 that restricts the estimation of $\eta_{j}$ to OLS and requires the second-stage covariates to strictly include the weighted average of shift-level covariates. We consider the $e_{n}$-weighted shift-share regression in line with \textcite{BorusyakHullJaravel2022}. $w_{j}$ denotes the shift weight $\sum_{i}e_{i}w_{ij}$ where, for a reminder, using $j$ over $j^{\prime}$ implies the range of the index: $j=1,\cdot\cdot\cdot,m$.

\noindent \textbf{Assumption D.3.} The $D_{j}$ are mutually independent given $\mathcal{G}_{m}$, $\max_j w_{j}\rightarrow0,$ and $\max \frac{w_{j}^{2}}{\sum_{j^{\prime}}w_{j^{\prime}}^{2}}\rightarrow0$.

\noindent \textbf{Assumption D.4.} (i) $E[|\eta_{j}|^{4+v}|\mathcal{F}_{m}]$ is uniformly bounded for some $v>0$; (ii) $\sum_{i,j}e_{i}w_{ij}^{2}V[\eta_{j}|\mathcal{F}_{m}]\pi_{ij}\ne0$ almost surely and $\sum_{i,j}e_{i}w_{ij}^{2}V[\eta_{j}|\mathcal{F}_{m}]\rightarrow\pi$ for a positive constant $\pi$; (iii) The support of $\pi_{ij}$ is bounded, the fourth moments of $\epsilon_{i}$, $v_{i}$, $p_{i}$, $\eta_{j}$ exist and are uniformly bounded, and $\sum_{i}e_{i}p_{i}p_{i}^{\top}\rightarrow\Omega_{pp}$ for positive definite $\Omega_{pp}$; (iv) The coefficient $\gamma$ to the control vector $p_{i}$ is consistently estimated by $\hat{\gamma}=(\sum_{i}e_{i}p_{i}p_{i}^{\top})^{-1}\sum_{i}e_{i}p_{i}\epsilon_{i}$ for true second-stage error $\epsilon_{i}$.

Finally, Assumption B4 is replaced with the consistency and regularity conditions on the shift residual estimator. This eliminates the need to include unit-level aggregate controls $\sum_{j}w_{ij}\mu_{j}$ even when $f$ is linear.

\noindent \textbf{Assumption D.5.} $(\hat{\eta}_{j})_{j}$ is a uniformly consistent estimator of $(\eta_{j})_{j}:||\hat{\eta}_{j}-\eta_{j}||_{2}\rightarrow0$.

We are ready to present the main result. Define $\hat{\beta}$ as the 2SLS estimator from the $e_{n}$-weighted shift-share regression of $Y_{i}$ on $X_{i}$ and $p_{i}$ instrumented with $Z_{i}=\sum_{j}w_{ij}\hat{\eta}_{j}$: $\hat{\beta}=\hat{E}[e_{i}Z_{i}Y_{i}]/\hat{E}[e_{i}Z_{i}X_{i}^{\perp}]$ where $\sum_{i}e_{i}=1$ and $X_{i}^{\perp}$ is the residual in the regression of $X_{i}$ on controls $p_{i}$.

\noindent \textbf{Proposition D.1.} Under Assumptions D.1-D.5, for $r_{m}=1/(\sum_{j}w_{j}^{2})$ and $\mathcal{V}=\text{plim}_{m\rightarrow\infty}r_{m}\mathcal{V}_{m}$:
\begin{equation*}
\sqrt{r_{m}}(\hat{\beta}-\beta) \xrightarrow{d} N(0, \frac{\mathcal{V}}{\pi^2})
\end{equation*}
where $\mathcal{V}_{m}=\sum_{j}(\sum_{i}e_{i}w_{ij}\epsilon_{i})^{2}V[\eta_{j}|\mathcal{F}_{m}]$. Furthermore,
\begin{equation*}
\frac{\sqrt{\sum_{j}(\sum_{i}e_{i}w_{ij}\hat{\epsilon}_{i})^{2}\hat{\eta}_{j}^{2}}}{|\sum_{i}e_{i}X_{i}^{\perp}Z_{i}|}
\end{equation*}
is an asymptotically valid standard error estimator for $\hat{\epsilon}$ the residual from the shift-share regression.

\noindent \textbf{Proof.} Define $\tilde{\beta}$ as the 2SLS estimator from the same regression but the instrument exploiting true mean-zero independent shocks rather than the estimated shocks: $\tilde{Z}_{i}=\sum_{j}w_{ij}\eta_{j}$. By equation (B22) of \textcite{BorusyakHullJaravel2022}, under Assumptions D.1-D.4, $\sqrt{r_{m}}(\tilde{\beta}-\beta)\rightarrow N(0, \frac{\mathcal{V}}{\pi^2})$. It suffices to show that $\sqrt{r_{m}}(\hat{\beta}-\tilde{\beta})=o_{p}(1)$. Next, for SE estimation:
\begin{align}
r_{m}(\sum_{j}(\sum_{i}e_{i}w_{ij}\hat{\epsilon}_{i})^{2}\hat{\eta}_{j}^{2}-\mathcal{V}_{m}) &= r_{m}(\sum_{j}(\sum_{i}e_{i}w_{ij}\epsilon_{i})^{2}\eta_{j}^{2}-\mathcal{V}_{m}) \tag{D.1} \\
&+ r_{m}(\sum_{j}(\sum_{i}e_{i}w_{ij}\hat{\epsilon}_{i})^{2}-(\sum_{i}e_{i}w_{ij}\epsilon_{i})^{2})\eta_{j}^{2} \tag{D.2} \\
&+ r_{m}\sum_{j}(\sum_{i}e_{i}w_{ij}\hat{\epsilon}_{i})^{2}(\hat{\eta}_{j}^{2}-\eta_{j}^{2}) \tag{D.3}
\end{align}
The first two terms are $o_{p}(1)$. The third term is also $o_{p}(1)$ since $r_{m}\sum_{j}(\sum_{i}w_{ij}\hat{\epsilon}_{i})^{2}$ is $O_{p}(1)$ and $\hat{\eta}_{j}^{2}-\eta_{j}^{2}\rightarrow0$ by Assumption D.5.

Setting the residual estimator $\hat{\eta}_{j}$ as that from the $w_{j}$-weighted OLS regression of $D_{j}$ on shift-level controls $q_{j}$ gives the heteroskedasticity-robust standard error in the reverted regression advocated in Borusyak, Hull and Jaravel (2022). The choice of estimator does not matter asymptotically as long as consistent. However, the choice may matter in finite samples. In \textcite{colantone2018a}, shifts are included only when there was an election. If the relation holds regardless, $\hat{\eta}_{j}$ can be more reliably estimated including all shifts.

\noindent \textbf{Assumption D.6.} $D_{j}\perp D_{j^{\prime}}$ if $c(j)\ne c(j^{\prime})$ given $\mathcal{G}_{m}$, $\max_c \tilde{w}_{c}\rightarrow0$.

\noindent \textbf{Proposition D.2.} Under Assumptions D.1-D.2 and D.4-D.6:
\begin{equation*}
\frac{\sqrt{\sum_{c}(\sum_{i}e_{i} \sum_{j:c(j)=c} w_{ij}\hat{\eta}_{j}\hat{\epsilon}_{i})^{2}}}{|\sum_{i}e_{i}X_{i}^{\perp}Z_{i}|}
\end{equation*}
is an asymptotically valid standard error estimator.

For the first-stage F-statistics, I propose using the asymptotic effective F-statistics in the inverted regression. Let us write the true first-stage regression as $\overline{X}_{j}^{\perp}=\alpha+\beta D_{j}+\gamma q_{j}+\overline{v}_{j}$. The effective F-statistics by \textcite{OleaPflueger2013} is equivalent to:
\begin{equation*}
\frac{E_{w}[(\alpha+\beta D_{j})^{2}]}{\text{tr}(\Sigma_{vv}\times\Sigma_{zz})}
\end{equation*}

From Frisch-Waugh-Lovell theorem with null $\gamma=0$ imposed, $\Sigma_{vv}$ can be estimated with the variance-covariance matrix of the regression $\overline{X}_{j}^{\perp}=\alpha+\beta\hat{\eta}_{j}+\overline{v}_{j}$.

\noindent \textbf{Proposition D.3.} The first-stage F-statistics can be consistently estimated with:
\begin{equation*}
\frac{\sum_{j}w_{j}(\hat{\overline{X}}_{j}^{\perp})^{2}}{\hat{\sigma}_{\beta\beta}\sum_{j}w_{j}D_{j}^{2}+2\hat{\sigma}_{\alpha\beta}\sum_{j}w_{j}D_{j}+\hat{\sigma}_{\alpha\alpha}\sum_{j}w_{j}}
\end{equation*}
where $\hat{\overline{X}}_{j}^{\perp}$ is the fitted value and $\hat{\sigma}_{ij}$ is the estimated covariance in regression (D.4).

% --- PAGE 48: Continuation of Proofs ---
is $O_{p}(1)$ by equation (B25) with $\tilde{g}$ replaced by 1, and $\hat{\eta}_{j}^{2}-\eta_{j}^{2}\rightarrow0$ by Assumption D.5. Setting the residual estimator $\hat{\eta}_{j}$ as that from the $w_{j}$-weighted OLS regression of $D_{j}$ on shift-level controls $q_{j}$ or $\hat{\eta}_{j}=D_{j}-(\sum_{j}w_{j}q_{j}q_{j}^{T})^{-1}(\sum_{j}w_{j}q_{j}D_{j})$, gives the heteroskedasticity-robust standard error in the reverted regression advocated in \textcite{BorusyakHullJaravel2022}. The above proposition shows that the choice of estimator does not matter asymptotically as long as consistent. However, the choice of estimator may matter in finite samples. In \textcite{colantone2018a}, shifts are included in the shift-share regression only when there was an election in the country in the given year. If we believe that relation (4) holds regardless of the election history, then $\hat{\eta}_{j}$ can be much more reliably estimated by including all shifts with no corresponding elections that took place.

\noindent \textbf{Assumption D.6.} $D_{j}\perp D_{j}^{\prime}$ if $c(j)\ne c(j^{\prime})$ given $g_{m}$ $max_{c}$ $\tilde{w}_{c}\rightarrow0$.

\noindent \textbf{Proposition D.2.} Under Assumptions D.1-D.2 and D.4-D.6,
\begin{equation*}
\frac{\sqrt{\sum_{c}(\sum_{i}e_{i} \sum_{j:c(j)=c} w_{ij}\hat{\eta}_{j}\hat{\epsilon}_{i})^{2}}}{|\sum_{i}e_{i}X_{i}^{\perp}Z_{i}|}
\end{equation*}
is an asymptotically valid standard error estimator.

% --- PAGE 50: F-Statistics ---
For the first-stage F-statistics, I propose using the asymptotic effective F-statistics in the inverted regression for the estimand, and estimating it with $\hat{\eta}_{j}$ where appropriate. Let us write the true first-stage regression as:
\begin{equation*}
\overline{X}_{j}^{\perp}=\alpha+\beta D_{j}+\gamma q_{j}+\overline{v}_{j} \tag{D.4}
\end{equation*}

\noindent \textbf{Proposition D.3.} The first-stage F-statistics can be consistently estimated with
\begin{equation*}
\frac{\sum_{j}w_{j}(\hat{\overline{X}}_{j}^{\perp})^{2}}{\hat{\sigma}_{\beta\beta}\sum_{j}w_{j}D_{j}^{2}+2\hat{\sigma}_{\alpha\beta}\sum_{j}w_{j}D_{j}+\hat{\sigma}_{\alpha\alpha}\sum_{j}w_{j}}
\end{equation*}
where $\hat{\overline{X}}_{j}^{\perp}$ is the fitted first-stage value and $\hat{\sigma}_{ij}$ is the estimated covariance between i and j in regression (D.4).

Note that $\hat{\eta}_j
= D_j-\left( \sum_j w_j q_j q_j^{\top} \right)^{-1} \left( \sum_j w_j q_j D_j \right)$
gives the first-stage F-statistics in \textcite{BorusyakHullJaravel2022}.

\newpage
% --- PAGE 51: Section E ---
\section*{E \quad Shift Replacement}
\textbf{Structural justification} \textcite{AutorDornHanson2013} derives the shift-share structure of the independent variable based on the monopolistic competition model where domestic production increases proportionally to imports. Shifts with small denominators represent cases where domestic production is likely being replaced by imports. Replacing outlier shifts with zero can be viewed as restricting the scope of analysis to industries that satisfy assumptions underlying the theoretical model.

\newpage

% --- PAGE 52: Appendix Tables ---
\section*{F \quad Appendix Figures and Tables}

\begin{table*}[htbp]
\centering
\caption*{Table F.1: Effect of China Shock on Electoral Outcomes per \textcite{colantone2018a}} 
\small
\begin{threeparttable}
\begin{tabular}{lcccccccccc}
\toprule
 & (1) & (2) & (3) & (4) & (5) & (6) & (7) & (8) & (9) & (10) \\
\cmidrule(lr){2-5} \cmidrule(lr){6-9} \cmidrule(lr){10-11}
 & \multicolumn{4}{c}{Nationalism} & \multicolumn{4}{c}{Nationalist Autarchy} & \multicolumn{2}{c}{Radical Right} \\
\cmidrule(lr){2-5} \cmidrule(lr){6-9} \cmidrule(lr){10-11}
Outcome: & \multicolumn{2}{c}{Median} & \multicolumn{2}{c}{COG} & \multicolumn{2}{c}{Median} & \multicolumn{2}{c}{COG} & \multicolumn{2}{c}{Share} \\
\midrule

\makecell[l]{Import\\Shock}
& 0.78$^{**}$ & 1.31$^{**}$ & 0.4$^{**}$ & 0.75$^{***}$
& 0.63$^{**}$ & 1.3$^{**}$ & 0.38$^{**}$ & 0.9$^{***}$
& 0.04$^{*}$ & 0.13$^{**}$ \\
& (0.33) & (0.47) & (0.15) & (0.22)
& (0.26) & (0.47) & (0.14) & (0.25)
& (0.02) & (0.05) \\
Obs & 8181 & 7782 & 8181 & 7782 & 8181 & 7782 & 8181 & 7782 & 8181 & 7782 \\

\addlinespace
\makecell[l]{Censored\\Shock}
& 0.43$^{*}$ & 0.53 & 0.28$^{**}$ & 0.36$^{**}$
& 0.43$^{*}$ & 0.65$^{**}$ & 0.32$^{**}$ & 0.67$^{***}$
& 0.01 & 0.03$^{**}$ \\
& (0.24) & (0.33) & (0.13) & (0.15)
& (0.22) & (0.28) & (0.13) & (0.20)
& (0.01) & (0.01) \\
Obs & 2757 & 2757 & 2757 & 2757 & 2757 & 2757 & 2757 & 2757 & 2757 & 2757 \\

\addlinespace
\makecell[l]{Imputed\\Shock}
& 0.63$^{**}$ & 0.77$^{**}$ & 0.31$^{**}$ & 0.33$^{***}$
& 0.32 & 0.51$^{***}$ & 0.26$^{**}$ & 0.42$^{***}$
& 0 & 0.01 \\
& (0.24) & (0.28) & (0.11) & (0.11)
& (0.22) & (0.17) & (0.09) & (0.10)
& (0.01) & (0.01) \\
Obs & 2761 & 2761 & 2761 & 2761 & 2761 & 2761 & 2761 & 2761 & 2761 & 2761 \\

\addlinespace
\makecell[l]{Predicted\\Shock}
& 0.67$^{**}$ & 0.76$^{**}$ & 0.34$^{***}$ & 0.33$^{***}$
& 0.32 & 0.49$^{**}$ & 0.26$^{**}$ & 0.41$^{***}$
& 0 & 0.01 \\
& (0.24) & (0.29) & (0.11) & (0.12)
& (0.23) & (0.17) & (0.09) & (0.10)
& (0.01) & (0.01) \\
Obs & 2761 & 2761 & 2761 & 2761 & 2761 & 2761 & 2761 & 2761 & 2761 & 2761 \\

Estimator & OLS & 2SLS & OLS & 2SLS & OLS & 2SLS & OLS & 2SLS & OLS & 2SLS \\

\bottomrule
\end{tabular}

\begin{tablenotes}[flushleft]

\footnotesize
\item
\textit{Note:} The first row replicates the results from the original table. The second row restricts the temporal scope to 2001 through 2007 for which reconstructed variables are available. The third row uses reconstructed variables, with missing employment shares imputed using linear regression. The fourth row replaces all observed values with predictions from linear regression. COG stands for the center of gravity, or the weighted average described in the section. All specifications have country-year fixed effects, and parentheses are standard errors clustered by NUTS-2 region-year. \hfill $^{*}p<0.1;\ ^{**}p<0.05;\ ^{***}p<0.01$
\end{tablenotes}

\end{threeparttable}
\end{table*}

\newpage

\begin{table*}[htbp]  
\caption*{Table F.2: Shift- and Unit-level Placebo Test With Raw shifts} 
\centering
\vspace*{\fill}         % push down

\small
\begin{threeparttable}
\label{tab:appendix:placebo-raw-shifts}

\begin{tabular}{p{0.62\textwidth}ccc}
\toprule
\textbf{Variable} & \textbf{Estimate} & \textbf{SE} & \textbf{Obs} \\
\midrule
shift-level: & & & \\
\quad Initial \% of national industry employment & $-0.227^{**}$ & (0.114) & 1370 \\
\addlinespace
Unit-level: & & & \\
\quad Initial \% of foreign-born population & 0.008 & (0.015) & 321 \\
\quad Initial \% of high-skilled workers & 0.658 & (0.999) & 335 \\
\quad Initial \% of high-technology workers & 0.121 & (0.248) & 335 \\
\quad Initial \% of medium- or low-skilled workers & $-0.062$ & (0.469) & 335 \\
\quad Initial \% of medium- or low-technology workers & 1.181 & (1.459) & 335 \\
\quad Initial \% of workers in primary sectors & $-1.265$ & (1.334) & 335 \\
\quad Initial \% of service industry workers & 2.683 & (3.529) & 335 \\
\bottomrule
\end{tabular}

\begin{tablenotes}[flushleft]
\footnotesize
\item
\textit{Note:} Specifications are all the same with Table 4 except that raw shifts were used instead of transformed shifts. The transformation only affects the shift-level test as most transformed shifts have too small shares to affect the shift-share instrument variable at the regional level.
\hfill $^{*}p<0.1;\ ^{**}p<0.05;\ ^{***}p<0.01$
\end{tablenotes}
\end{threeparttable}
\vspace*{\fill}         % push up (centers vertically)
\end{table*}

\newpage

\begin{table*}[p]
\centering
\caption*{Table F.3: Table 6 with Predicted Shares and Untransformed Shifts}
\label{tab:appendix:table6-predshares-rawshifts}
\small
\begin{threeparttable}
\begin{tabular}{lcccccc}
\toprule
 & (1) & (2) & (3) & (4) & (5) & (6) \\
\cmidrule(lr){2-4}\cmidrule(lr){5-7}
 & \multicolumn{3}{c}{Median} & \multicolumn{3}{c}{Center of Gravity} \\
\midrule

\multicolumn{7}{l}{Predicted Shares:} \\
\addlinespace
\quad Cluster
& 0.745$^{**}$ & 0.683$^{**}$ & $-0.455$
& 0.301$^{***}$ & 0.321$^{***}$ & $-0.224$ \\
& (0.292) & (0.281) & (0.349)
& (0.115) & (0.119) & (0.147) \\
\quad Obs
& 3006 & 307 & 307
& 3006 & 307 & 307 \\
\quad F
& 1375.02 & 174.08 & 33.42
& 1375.02 & 174.08 & 33.42 \\
\addlinespace
\quad BHJ
& -- & 0.683$^{***}$ & $-0.455$
& -- & 0.321$^{***}$ & $-0.224$ \\
& -- & (0.205) & (0.588)
& -- & (0.094) & (0.49) \\
\quad Obs
& -- & 349 & 349
& -- & 349 & 349 \\
\quad F
& -- & 68.17 & 1.55
& -- & 68.17 & 1.55 \\

\addlinespace\addlinespace
\multicolumn{7}{l}{Untransformed shifts:} \\
\addlinespace
\quad Cluster
& 0.77$^{***}$ & 0.629$^{***}$ & $-0.023$
& 0.334$^{***}$ & 0.312$^{***}$ & $-0.15$ \\
& (0.28) & (0.242) & (0.659)
& (0.113) & (0.108) & (0.209) \\
\quad Obs
& 3006 & 307 & 307
& 3006 & 307 & 307 \\
\quad F
& 1546.76 & 174.15 & 125.87
& 1546.76 & 174.15 & 125.87 \\
\addlinespace
\quad BHJ
& -- & 0.629$^{***}$ & $-0.023$
& -- & 0.312$^{***}$ & $-0.15$ \\
& -- & (0.177) & (0.892)
& -- & (0.084) & (0.21) \\
\quad Obs
& -- & 349 & 349
& -- & 349 & 349 \\
\quad F
& -- & 46.82 & 1
& -- & 46.82 & 1 \\

\addlinespace\addlinespace
\multicolumn{7}{l}{Predicted Shares and untransformed shifts:} \\
\addlinespace
\quad Cluster
& 0.756$^{***}$ & 0.667$^{**}$ & $-0.442$
& 0.327$^{***}$ & 0.323$^{***}$ & $-0.233^{*}$ \\
& (0.291) & (0.265) & (0.302)
& (0.115) & (0.114) & (0.134) \\
\quad Obs
& 3006 & 307 & 307
& 3006 & 307 & 307 \\
\quad F
& 1414.12 & 172.93 & 220.19
& 1414.12 & 172.93 & 220.19 \\
\addlinespace
\quad BHJ
& -- & 0.667$^{***}$ & $-0.442$
& -- & 0.323$^{***}$ & $-0.233$ \\
& -- & (0.179) & (0.494)
& -- & (0.083) & (0.416) \\
\quad Obs
& -- & 349 & 349
& -- & 349 & 349 \\
\quad F
& -- & 39.63 & 1.01
& -- & 39.63 & 1.01 \\

\addlinespace\addlinespace
\quad Unit of Analysis
& District & Region & Region
& District & Region & Region \\
\quad shift controls
& F & F & T
& F & F & T \\
\quad Country-Year FE
& T & T & T
& T & T & T \\

\bottomrule
\end{tabular}

\begin{tablenotes}[flushleft]
\footnotesize
\item
\textit{Note:} All specifications are the same with Table 6 besides shifts and shares. \\
\hfill $^{*}p<0.1;\ ^{**}p<0.05;\ ^{***}p<0.01$
\end{tablenotes}
\end{threeparttable}
\end{table*}

\clearpage        % finish current page
\FloatBarrier    
\end{document}